\documentclass[prb,twocolumn,showpacs]{revtex4-1}

\usepackage{graphicx,epstopdf}
\usepackage{amsmath}
\usepackage{amssymb}
\usepackage{color}
\usepackage{float}
\usepackage{epsfig}
\usepackage{epstopdf}
\usepackage{enumitem}

\newcommand\T{\rule{0pt}{2.4ex}}
\newcommand\B{\rule[-1.2ex]{0pt}{0pt}}
\usepackage[colorlinks=true, citecolor=blue, urlcolor=blue ]{hyperref}
\input{epsf}
\begin{document}

%\title{Genuine multipartite entanglement in resonating valence bond states under doping}
\title{Analytical recursive method to ascertain multisite entanglement \\ in doped quantum spin ladders }
%\title{Genuine multipartite entanglement in the superconducting phase \\of doped resonating valence bond ladders}
%\title{Genuine multipartite entanglement in superconducting \textit{t-J} ladders}

\author{Sudipto Singha Roy$^{1,2}$, Himadri Shekhar Dhar$^{1,2,3}$, Debraj Rakshit$^{1,2,4}$, Aditi Sen(De)$^{1,2}$, and Ujjwal Sen$^{1,2}$}

\affiliation{$^1$Harish-Chandra Research Institute, Chhatnag Road, Jhunsi, Allahabad 211 019, India \\
$^2$Homi Bhabha National Institute, Training School Complex, Anushakti Nagar, Mumbai 400 085, India\\
$^3$Institute for Theoretical Physics, Vienna University of Technology, Wiedner Hauptstraße 8-10/136, A-1040 Vienna, Austria\\ $^4$Institute of Physics, Polish Academy of Sciences, Aleja Lotnik\'ow 32/46, PL-02668 Warsaw, Poland}

\date{\today}

\begin{abstract}

%The high-temperature superconducting phases of cuprates and pnictides have been largely described by doping the Mott insulator.
%In our work,
We formulate an analytical recursive method to generate the wave function of doped short-range resonating valence bond (RVB) states
%, which are potential ground states of a Hubbard model with large onsite interactions,
as a tool to efficiently estimate multisite entanglement as well as other physical quantities in doped quantum spin ladders. We prove that doped RVB ladder states are always genuine multipartite entangled. Importantly, our results show that within  specific doping concentration and model parameter regimes, the doped RVB state essentially characterizes the trends of genuine multiparty entanglement in the exact ground states of the Hubbard model with large onsite interactions,  in the limit which yields the $t$-$J$ Hamiltonian.

\end{abstract}
\maketitle

\section{Introduction}
%\emph{Introduction.}--
%\textcolor{red}{
One of the best known and simplest theoretical frameworks for investigating
{ strongly-correlated doped quantum spin ladders is the $t$-$J$ model, which is obtained in the limit of large on-site interaction from the Hubbard model~\cite{t-J,comm-tJ,t-J2,t-Janti}. At half-filling, without doping, the system reduces to a Heisenberg ladder with a spin liquid ground state (GS)~\cite{spinliquid}. Upon doping the spin ladder, studies based on mean-field theory using Gutzwiller renormalization show that the spin gap is persistent~\cite{mean} which is a tell-tale sign of strong superconducting fluctuations\cite{spinliquid,mean,telltale}.
%Most developments in high-$T_c$ superconductivity involve antiferromagnetic order in doped Mott insulators \cite{Mott} based on transition-metal oxides \cite{sc_3,sc_4}, as compared to the more conventional superconductivity \cite{onnes} at low temperatures based on the BCS theory \cite{bcs}
%(cf. \cite{comm0,h2s,h2s2}). Over the years, notable progress has been made in experimental investigation of high-$T_c$ superconductivity \cite{SC_exp} but the microscopic theory behind this novel phenomenon remains unresolved \cite{knowno}. One of the best known and simplest theoretical frameworks for investigating high-$T_c$ superconductivity is the $t$-$J$ ladder \cite{Anderson_tJ,tj}, which is obtained in the limit of
%large on-site interaction from the Hubbard model \cite{t-J}. At half-filling, without doping, the system reduces to a Heisenberg ladder with a spin liquid ground state (GS) \cite{spinliquid}. Upon doping the $t$-$J$ ladder, studies based on mean-field theory using Gutzwiller renormalization show that the spin gap is persistent \cite{mean}, which is a tell-tale sign of strong superconducting fluctuations \cite{spinliquid,mean,telltale}. The superconducting states of the $t$-$J$ ladder can be represented using the short-range resonating valence bond (RVB) ansatz \cite{anders_rvb, Anderson_RVB}, which were introduced to describe Mott-insulators %phase of the
%in spin-1/2 Heisenberg antiferromagnets (AFMs) \cite{anders_rvb}.\
The $t$-$J$ model, under finite doping, exhibits a rich phase diagram, which has been extensively studied for low-dimensional antiferromagnets (AFM)~\cite{t-J_phase,t-J_refernce, dimer-hole1, dimer-hole2}. In particular, in 1D and ladder configurations, the system possess exotic correlation properties that are characterized by the Luttinger liquid theory~\cite{ll}, as confirmed using exact diagonalization calculations, and exhibits  a rich superconducting phase for a  specific range of values of $J/t$ and electron density, $n_{el}$~\cite{t-J_phase,sc_3,sc_4,t-J_refernce, dimer-hole1, dimer-hole2}.
Moreover, the superconducting states of the quantum spin ladder can be represented using the short-range resonating valence bond (RVB) ansatz~\cite{anders_rvb, Anderson_RVB} which were introduced to describe Mott-insulators.\

%phase of the
%in spin-1/2 Heisenberg antiferromagnets (AFMs) \cite{anders_rvb}.\
% In this work, our interest lies in investigating $t$-$J$ ladders in the region $J/t \gtrsim 0.5$ where the superconducting phase seems to appear at relatively high $n_{el}$ \cite{t-J_phase}.
%This is also influenced by the fact that these $J/t$ regimes can potentially be realized in fermionic ultracold gases at high energy scales \cite{energy-scale}.
 An important yet demanding  task in the study of doped quantum spin ladders is to characterize how quantum correlations, in particular multiparty entanglement,  are distributed among the subparts of these strongly correlated systems. This is motivated, on one hand, by the fact that study of multisite physical quantities in many-body quantum systems often provide deeper insights into the cooperative phenomena they exhibit~\cite{fazio,sir_maam}. On the other hand, such investigation can play an important role  for implementation of quantum information processing tasks in the laboratory. However, estimation of the same remains a challenging task, primarily due to the exponential growth of the Hilbert space with increasing system size.
This is especially true if we try to obtain analytical expressions or bounds of multisite physical properties such as entanglement.
%{\color{red}This is especially  true if we try to obtain analytical expressions or
%bounds of the multisite physical properties like entanglement \cite{} of
%pure or mixed states.
Therefore, obtaining a general method to characterize
entanglement in multipartite states is crucial to investigate
physical phenomena of a complex system.

%}
%However, due to the exponential growth of the Hilbert space with the increase of particle numbers, even for the moderate-sized system, it becomes quite a challenging task to deal the system via exact diagonaliza-tion.

In this work, we consider short-range doped resonating valence bond (RVB) states,
% to study the multiparty entanglement properties of the GS of the $t$-$J$ model.
%
%
%
and for finite values of the electron density ($n_{el}$), using the symmetry properties of the RVB state~\cite{chan,ours3,ours1}, we prove that the doped RVB ladder is always genuinely multipartite entangled.  To quantify
the genuine multiparty entanglement %To obtain $\cal{G}$
in large spin ladders,
%, where exact calculations are inaccessible,
we introduce an analytical recursion method to build the doped RVB state. The novelty of this recursion method stems from the fact that in a large spin network with arbitrary electron density ($n_{el}$), one can analytically compute the reduced density matrices of the superposition state, thus allowing an exact estimation of the genuine multiparty entanglement using the generalized geometric measure ($\cal{G}$)\cite {ggm1} (cf.~Refs.\cite{mul,ggm2}).
%, for a range of
%doping concentration ($x$) or its complementary quantity,
%the electron density ($n_{el}$).
Using the proposed recursion method, we observe that in the thermodynamic limit, $\cal{G}$ increases with $n_{el}$, reaching a maximum at $n_c \approx 0.56$, before decreasing for higher $n_{el}$ (cf.~\cite{ravi}).
%
%We compare this behavior of genuine multipartite entanglement with those calculated for the GS of the doped $t$-$J$ ladder obtained through exact diagonalization for moderate-sized lattice.
%\textcolor{red}{
Interestingly, we further show that the qualitative multipartite features of doped RVB states are  closely mimicked by  ground states (GSs) of doped \(t\)-\(J\) ladders obtained through exact diagonalization for moderate-sized lattices. In particular, we present a representative case with $J/t \approx 0.6$, where we observe that genuine multiparty entanglement of the GS of the $t$-$J$ ladder emulates the same %similar behavior in terms of the genuine multiparty entanglement
of the doped RVB state.
%}
The maximum $\mathcal{G}$ occurs at $n_c \approx 0.65$, close to that obtained using the doped RVB ansatz.  The discrepancy in the values of electron densities  need to  account for finite size effect.
Hence, using the analytical recursion method, one can show that
%the maximum genuine multipartite entanglement in doped RVB ladders corresponds to the strongly-correlated superconducting phase of the $t$-$J$ model, and more importantly,
%the recursion method obtained here enables us to show that
within the considered
parameter range, the trend of genuine multipartite entanglement of the
former state qualitatively matches with that of the GS of the latter model.
%Importantly, our results show that, within the considered parameter range, the doped RVB state qualitatively characterizes the genuine multiparty entanglement properties of the GS of $t$-$J$ ladders.
%can essentially be characterized using the doped RVB ansatz}.
We note that although we use the recursion method to study multipartite entanglement, the method can also be employed to investigate other properties like single-site, two-site physical quantities of the doped RVB ladders for systems with an arbitrary number of sites. In our calculations, we have considered up to 300 sites, and though a higher number of sites are accessible through our method, the physical quantities of interest converge much earlier.

The paper is organized as follows. In Sec.~\ref{recursion_state}, we present the recursion method that  generates the doped RVB state  corresponding to  arbitrary electron density. Thereafter, in Sec~\ref{recursion_reduced}, we propose the recursion relation for density matrices of multiple sites, considering  both periodic as well as open boundary conditions. In Sec.~\ref{genuine}, we provide analytical results on  genuine multipartite entanglement in the doped RVB state. We  also discuss a measure  of genuine multipartite entanglement called the generalized geometric measure ($\mathcal{G}$) in Sec.\ref{ggm_definition}.
% to quantify the genuine multiparty entanglement.
In Sec.~\ref{model}, we introduce the $t$-$J$ model and subsequently compare the behavior of $\mathcal{G}$ obtained using the doped RVB state ansatz to that obtained via  exact diagonalization of the $t$-$J$ model. We present a brief conclusion  in Sec.~\ref{conclusion}.

\section{Recursion method to generate doped RVB states}
\label{recursion_state}
%\emph{Recursion method to generate doped RVB states.--}
We begin by considering
%
%Considering the short-range doped RVB state as a framework to study the GS entanglement properties of the $t$-$J$ model,
%we begin with
the doped RVB state containing $2N$ lattice sites, on a ladder configuration, with $2k$ spin-1/2 particles and $2(N-k)$ holes or vacant lattice sites, expressed
%To express the dimer coverings in an RVB state consistently, one represents the ladder
using a \textit{bipartite lattice}, consisting of sublattices $A$ and $B$.
%The dimer state is formed between a spin particle in
%The doped RVB state consists of equal-weight superposition of all possible coverings of $k$ NN dimers with $2(N-k)$ holes arranged in such a way as to allow a complete dimer covering.
%, i.e., no spin particles are left outside a dimer in the covering.
The corresponding (unnormalized) wavefunction, with electron density $n_{el}$ = $k/N$, is given by
%%
%%consisting of We consider a pure state configuration formed by equal superposition of all possible NN dimer coverings in a 2-legged, $N$-runged doped ladder system, conventionally known as doped RVB state. Therefore for a ladder comprised of $2N$ numbers of lattice sites and $N-k$ numbers of hole pairs the RVB state can be defined
%%in the following way, \
\begin{eqnarray}
|\Psi\rangle_{k,N-k}&=&\sum_i r_i |(a_{n_1} b_{n_1})(a_{n_2} b_{n_2})\cdots(a_{n_k} b_{n_k})\rangle_i \nonumber\\
&&\otimes~ |h_{m_{2k+1}}h_{m_{2k+2}}\cdots h_{m_{2N}}\rangle_i,
\label{Eqn:RVB}
\end{eqnarray}
where $|(a_{n_j} b_{n_j})\rangle=\frac{1}{\sqrt{2}} (|01\rangle-|10\rangle)_{n_j}$ represents a dimer, with $a_j \in A$ and $b_j \in B$. $|\{(a_{n_j} b_{n_j})\}\rangle_i$ represents a complete dimer covering at occupied sites $n_j$. The holes,
$|h_{m_j}\rangle=|2\rangle_{m_j}$, are at sites $m_j$, such that $\sum_{j=1}^{k} 2n_j$ + $\sum_{j=2k+1}^{2N} m_j$ = $2N$. $r_i$ = 1, $\forall~ i$.\\
%%The range of $i$ in the summation is equal to the total number of possible coverings, estimation of which is an \textit{NP}-complete problem \cite{ref20ved}.
%
%We note that the holes in the lattice are equally distributed between sublattices $A$ and $B$ to necessitate complete dimer coverings but can migrate across all the $2N$ lattice points. See the Appendix for further description.
%
%To overcome
%%As size of the ladder increases altogether with the hole-singlet ratio, the number of configurations of the dimer-hole coverings of the lattice increases exponentially.
%the complexity \cite{complex} in analyzing large doped RVB ladders, we recursively \cite{ours2,delgado,fan, delgado2} construct the state $|\psi\rangle_{k,N-k}$, defined in Eq.~(\ref{Eqn:RVB}) and generate its reduced states.
%\vspace{0.1cm}
%{\it Recursive method to generate the doped RVB state.}--

In general, in considering the RVB ansatz for the ground state of a moderate-sized doped quantum spin ladder, as described in Eq.~(\ref{Eqn:RVB}), the number of dimer coverings in the state increases exponentially with the increase of the electron density~\cite{complex}. For example, in a small spin ladder with 5 spins on each leg, the number of dimer coverings at electron density $n_{el}=0.33$  94, and at $n_{el}=0.66$, it is equal to 294. Hence, even for small ladders, a direct construction of the RVB ground state is computationally expensive. Moreover, the Hilbert space also increases rapidly with increase in the number of spins. This makes the analytical recursion method proposed for studying physical properties of doped RVB states on large quantum spin ladders, a very important part of our results. We recursively~\cite{ours2,delgado,fan,delgado2} construct the state $|\psi\rangle_{k,N-k}$, defined in Eq.~(\ref{Eqn:RVB}) and generate its reduced states. Though earlier attempts have been made to obtain recursion relations for physical observables such as the ground state energy~\cite{delgado}, the novelty of our approach lies in the fact that the proposed method recursively constructs the reduced density matrices of the doped RVB state, which allows us to study quantum and classical properties, in particular, multipartite entanglement, which in turn are used to characterize the system. \
%%
%%here we propose a recursion relation using which one can construct the doped RVB state $|\psi\rangle_{N-k,k}$ for arbitrary number of spins
%%and doping concentration.

In order to generate the analytical recursion method, let us begin with an open $2N$-site ladder lattice with all vacant sites (holes), which is successively filled with dimers. We use the notation, $|N-k,k\rangle$ to denote the $N$-rung ladder, $|\psi\rangle_{k,N-k}$, containing %arbitrary number of spins
$2k$ spins filled with dimers and $2(N-k)$ holes. The state $|N-k,k\rangle$ is achieved by successively filling $k$ dimers in the $|N,0\rangle$ state, %e.g., an 8 site RVB ladder doped with 4 holes, can be generated as: $|4,0\rangle$ $\xrightarrow{\mathcal{U}^{\otimes{k=1}}}$ $|3,1\rangle$ $\xrightarrow{\mathcal{U}^{\otimes{k=1}}}$ $|2,2\rangle$,
i.e., $|N,0\rangle$ $\xrightarrow{{k}}$ $|N-k,k\rangle$. As an example, consider an initial configuration with  8 site RVB ladder, doped with 4 holes. Now the state $|2,2\rangle$ mentioned above, can be generated in the following way:
\begin{equation}
|4,0\rangle \xrightarrow{{{~~k=1~~}}} |3,1\rangle \xrightarrow{{{~~k=1~~}}} |2,2\rangle,\nonumber
\label{example}
\end{equation}
where $|4,0\rangle$ is the initial lattice with all holes, and $|2,2\rangle$ is the final state, for an 8 site RVB ladder, with $2$ dimers and $2$ pairs of holes.\

For an analytical method which allows us to build the superpositions in an arbitrary $|N-k,k\rangle$, we propose the generator% using the relation:
\begin{eqnarray}
&&|N-k,k\rangle = \mathcal{U}^{\otimes{k'=1}}|N-k+1,k-1\rangle \nonumber\\&+&
|N-k-1,0\rangle |\chi_{k+1}\rangle +
|N-k-2,0\rangle |\chi_{k+1} \rangle |1,0\rangle,\nonumber\\
\label{eq:generator}
\end{eqnarray}
where  $\mathcal{U}^{\otimes{k'}}$ is the operator to add $k'$ dimers.
The methodology to derive the above recursion relation and the description of $|\chi_{k+1} \rangle$ are given below.
%Let us begin with a $2N$ ladder lattice with all vacant sites (holes), which is successively filled with dimers. We use the notation, $|N-k,k\rangle$ to denote the $N$-rung ladder state, $|\psi\rangle_{k,N-k}$, covered by $k$ dimers and $2(N-k)$ holes. The state $|N-k,k\rangle$ is achieved by successively filling $k$ dimers in the $|N,0\rangle$ state, which is the initial completely separable state containing only holes.
%The above scheme can be demonstrated through the following example.

%%where  $\mathcal{U}^{\otimes{k}}$ is the operator to fill $k$ dimers.
%To allow for an analytical approach to build the superposition in an arbitrary $|N-k,k\rangle$, we formulate a recursion method based on the above scheme.

%An important aim of the recursion method to generate the doped RVB state $|N-k,k\rangle$ is to successfully obtain the reduced state $\rho_{red}$.
To facilitate our calculations, we divide the $2N$ ladder lattice into specific regions that can be filled with dimers. We start by
%For a $m$-runged configuration where initially all the lattice sites are comprised with holes, the doped RVB state can be
splitting the initial state $|N,0\rangle$ into two regions, denoted by left ($L$) and right ($R$) block, such that
\begin{equation}
|N,0\rangle = |N-2,0\rangle_{L}~~\otimes~~|2,0\rangle_{R}.
\end{equation}

\begin{figure}[h]
\includegraphics[angle=0,width=8.5cm]{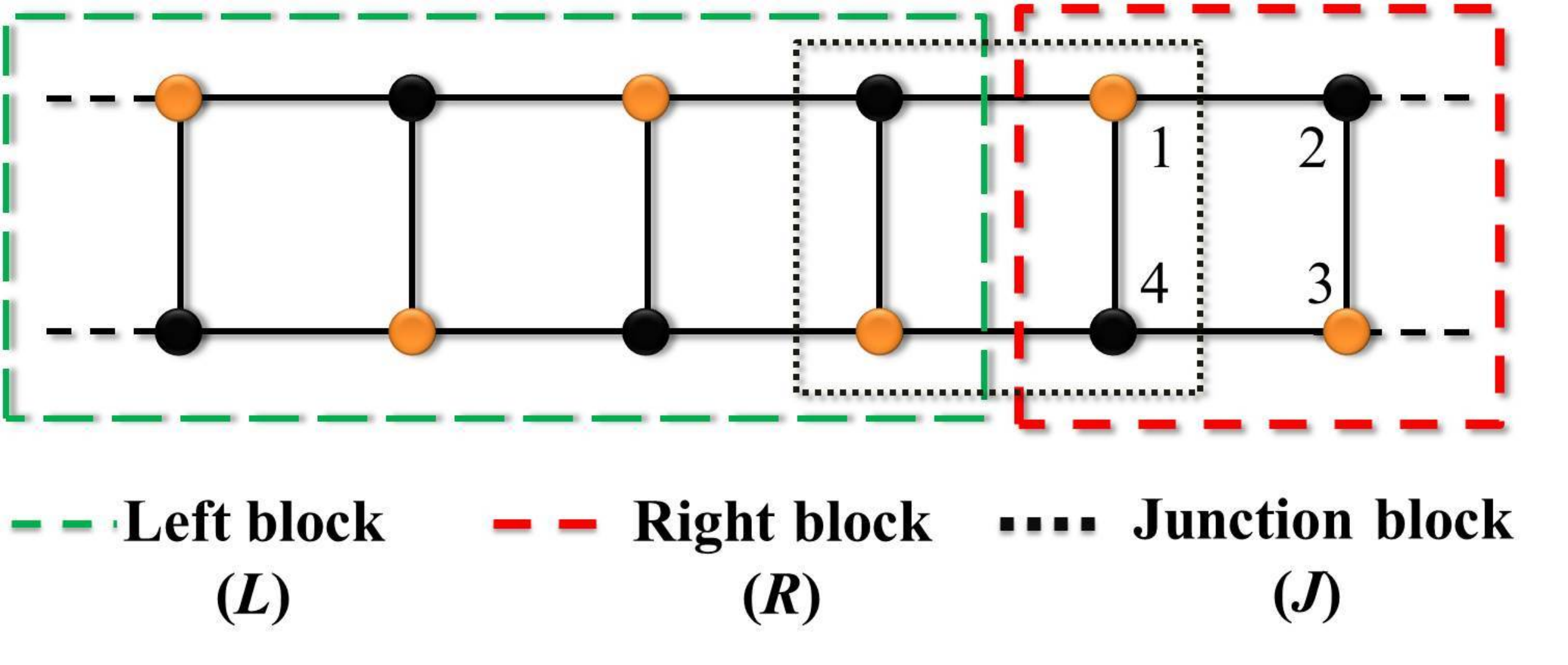}
\caption{Schematic diagram of the blocks $L$, $R$, and $J$ in the spin lattice. To compute $\mathcal{G}$, we obtain the reduced density matrix ($\rho_{red}$) corresponding to the sites 1-4 in the $R$ block. The rest of the lattice is traced out. Numerical studies show that the reduced state $\rho_{red}$ is sufficient to compute $\mathcal{G}$  in doped RVB states.  }
\label{fig2h}
\end{figure}

This is explicitly shown in Fig.~\ref{fig2h}.
%We note that the block $R$ is the region which contain the desired reduced density matrix ($\rho_{red}$), obtained by tracing out the state on the rest of the lattice sites.
An important region is the junction ($J$) block between $L$ and $R$ blocks, which is shown in Fig.~\ref{fig2h} using a black-dotted square. The blocks, excluding overlapping region,
can be written as:
%there are sites at the junction of these two bulk parts and those can be expressed as
\begin{equation}
|N-3,0\rangle_{L'}~~\otimes~~|2,0\rangle_{J}~~\otimes~~|1,0\rangle_{R'},\nonumber
\end{equation}
where $L'$ ($R'$) implies the region $L - L \cap J$ ($R - R \cap J$).

\begin{figure}
\begin{center}
\includegraphics[angle=0,width=3.6cm]{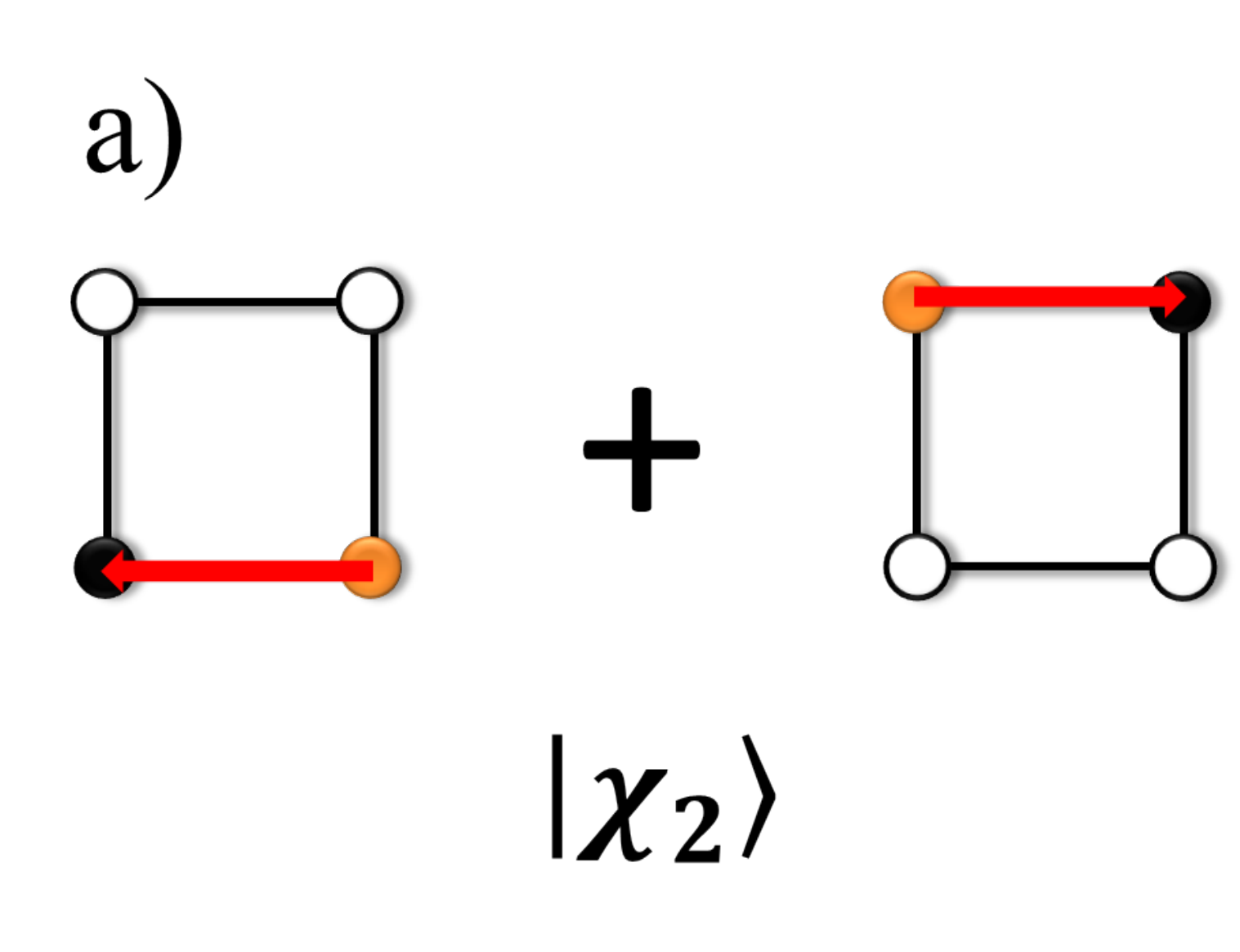}~~~
\includegraphics[angle=0,width=4.3cm]{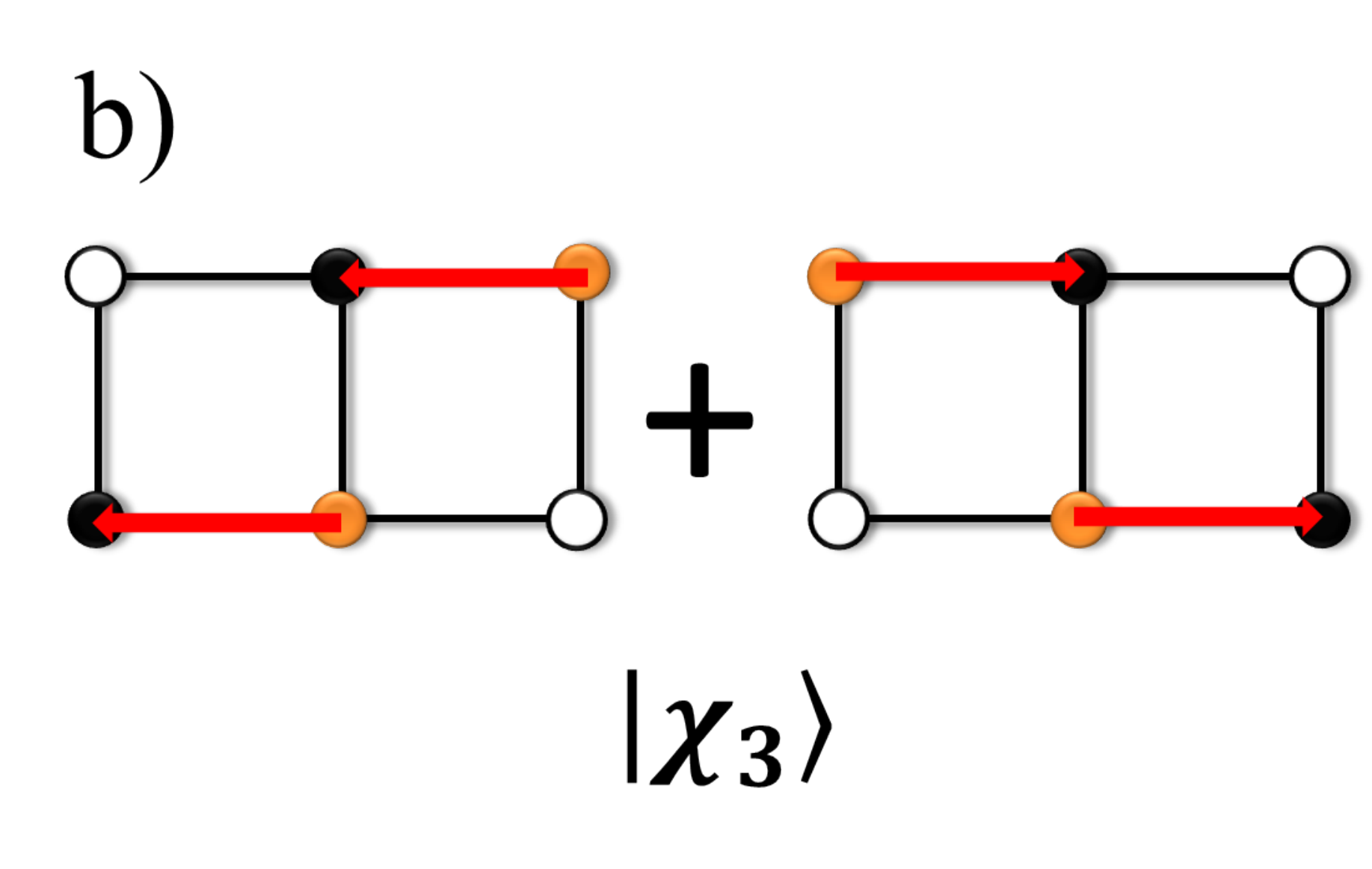}\\
\includegraphics[angle=0,width=6cm]{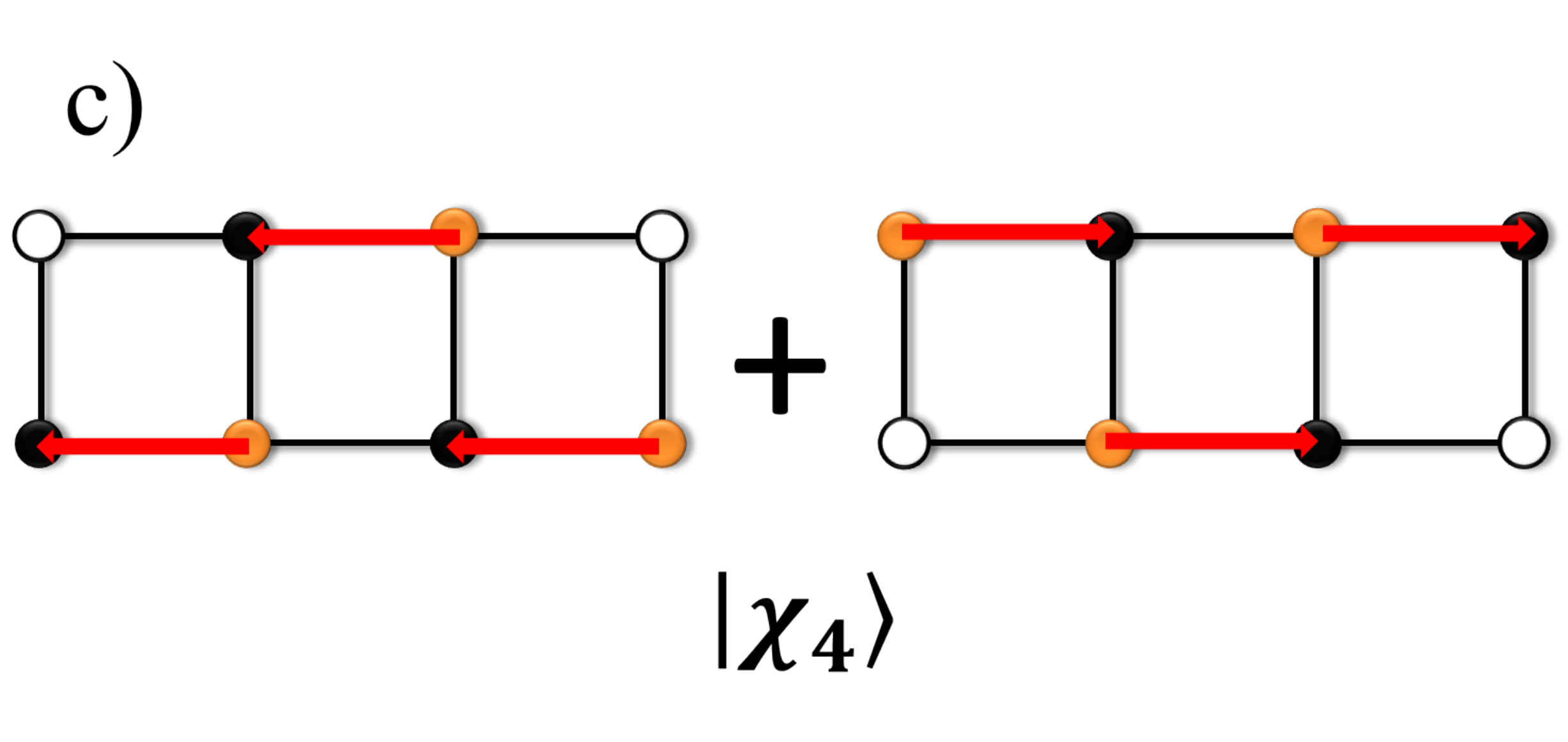}\\
\includegraphics[angle=0,width=6cm]{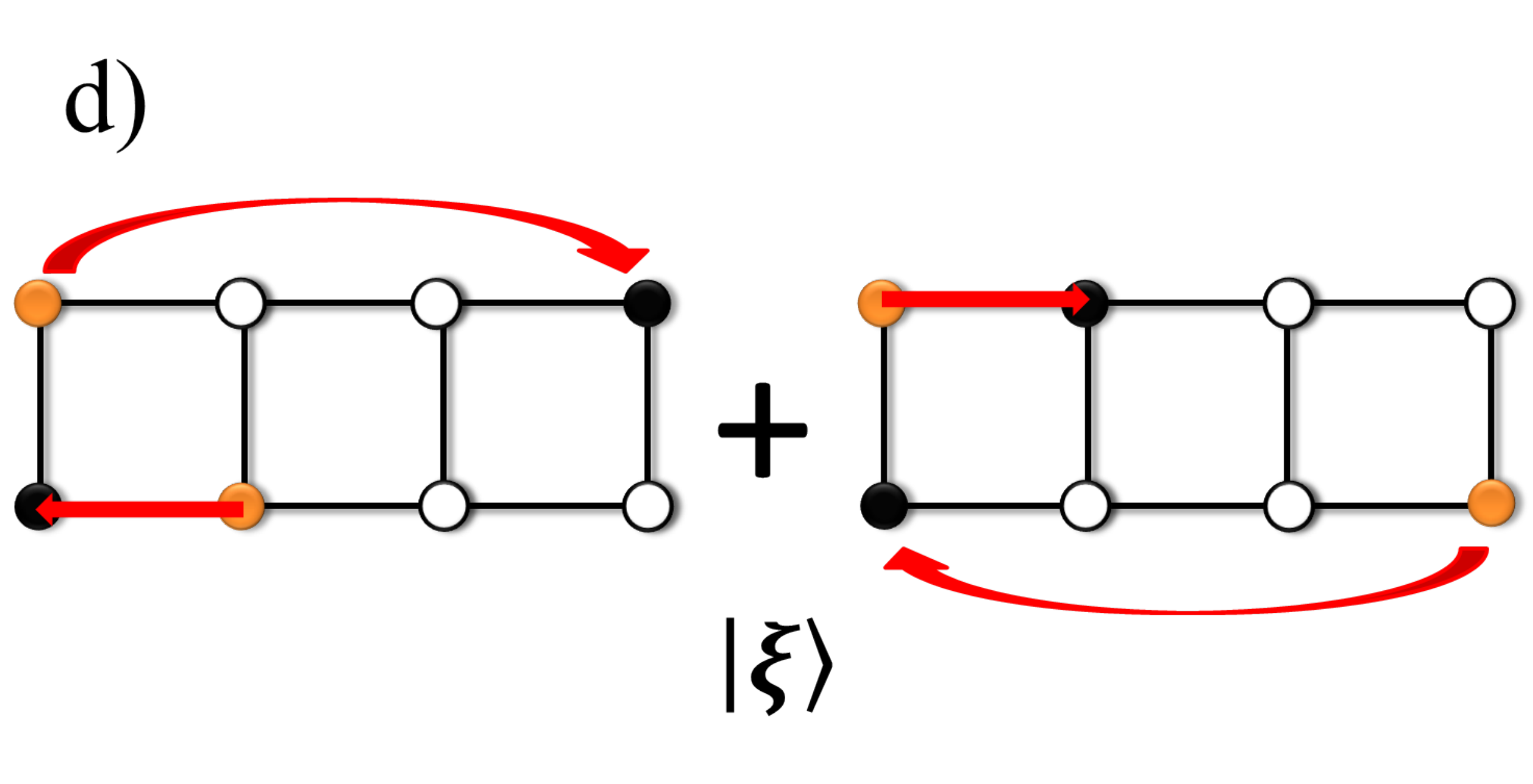} \\
\includegraphics[angle=0,width=6cm]{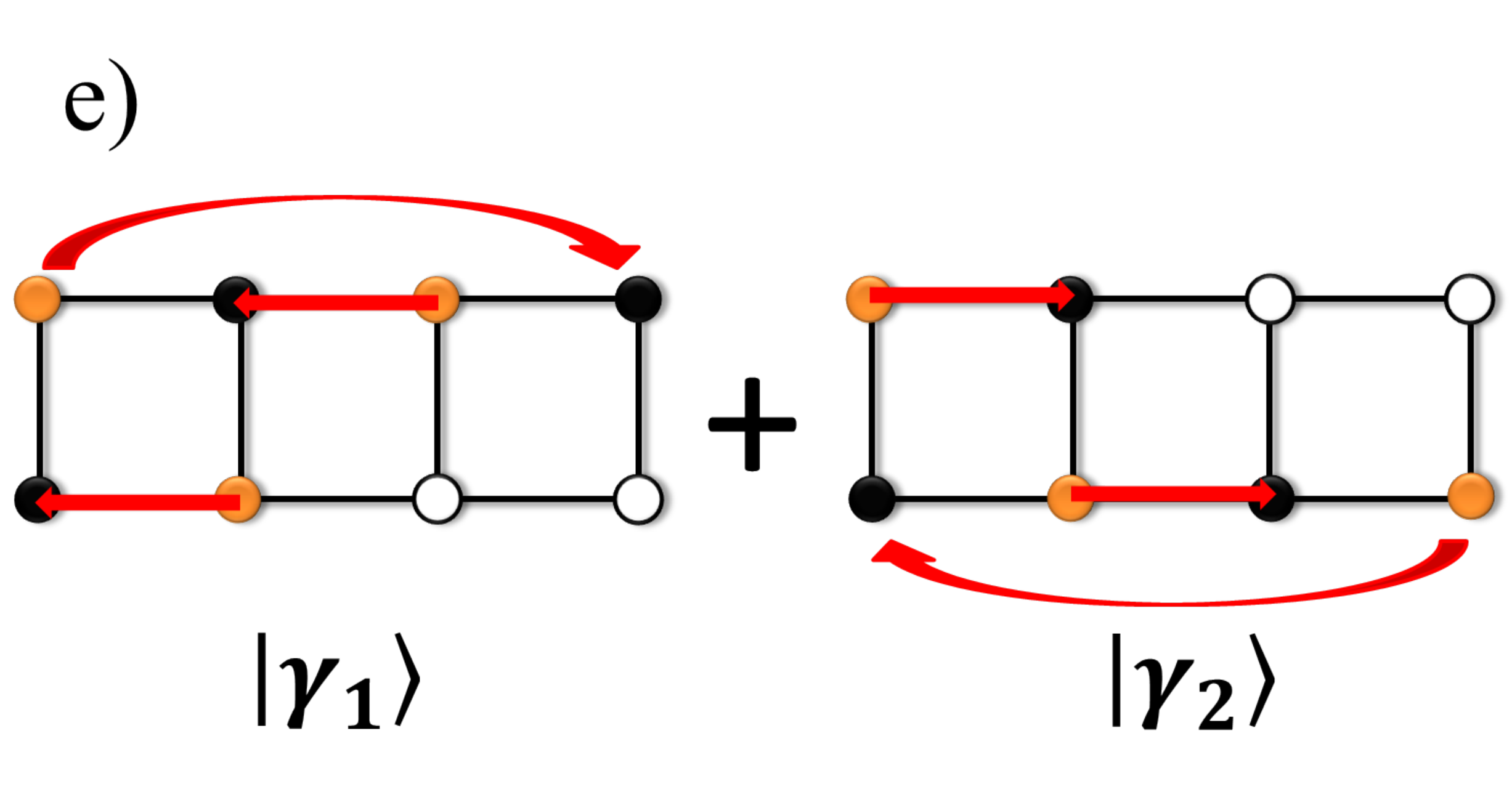}
\caption{Schematic diagram of the states: a) $|\chi_2\rangle$, b) $|\chi_3\rangle$, c) $|\chi_3\rangle$, and the periodic terms: d)~$|\xi\rangle$, and e)~$|\gamma_1\rangle$ and
$|\gamma_2\rangle$, used in the recursion relations.}
\label{fig3h}
\end{center}
\end{figure}

%\vspace{-.2cm}
Now starting from an initial configuration $|N,0\rangle$, our aim is to reach the final state $|N-k,k\rangle$ by systematically introducing $k$ numbers of dimers in the different blocks of the lattice.
%, we need to introduce the singlets systematically.
The first dimer is introduced in the initial hole configuration though the following possible ways:

\begin{enumerate}[leftmargin=0cm,itemindent=.5cm,labelwidth=\itemindent,labelsep=0cm,align=left]
   \begin{figure*}[t]
\includegraphics[angle=0,width=12cm]{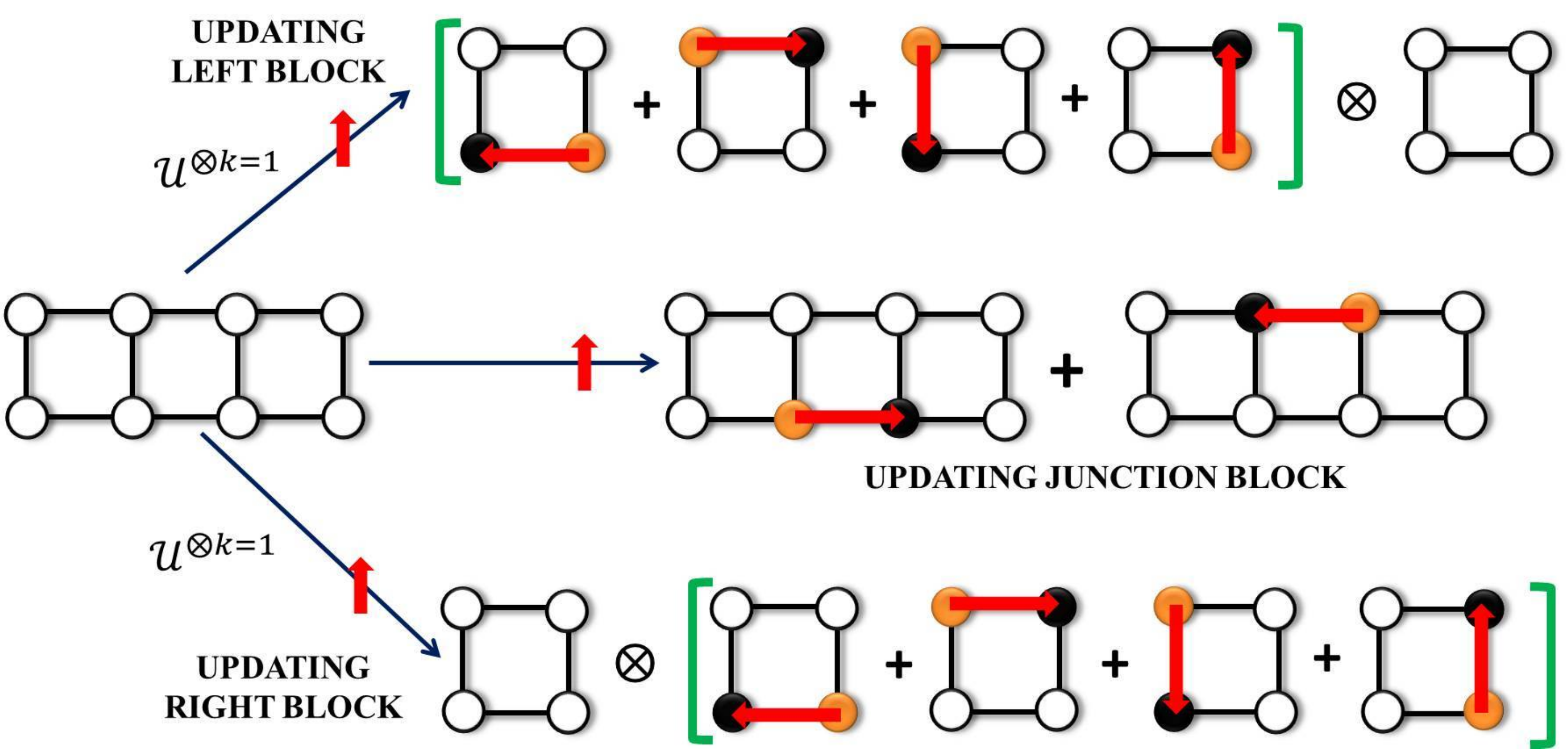}
\caption{Schematic diagram of the scheme to build a doped-RVB state from a lattice filled with holes, for an 8 site RVB ladder, as given by the process: $|4,0\rangle \xrightarrow{~~k=1~~}|3,1\rangle$, described in Eq.~(\ref{example}).}
\label{fig4h}
\end{figure*}

\item {\it{Update the left block}}: In this step, a dimer is introduced into the $L$ block  and the updated state is
\begin{equation}
|N-2,0\rangle_{L}~|2,0\rangle_{R}\xrightarrow{k=1} |N-3,1\rangle_{L}~|2,0\rangle_{R}.
\end{equation}
 \item {\it {Update the right block}}: Similarly, in the next step, a dimer is injected into the $R$ block. The updated state looks like
\begin{equation}
|N-2,0\rangle_{L}~ |2,0\rangle_{R}\xrightarrow{k=1} |N-2,0\rangle_{L} ~|1,1\rangle_{R}.
\end{equation}

\item {\it {Update of the junction block}}: In this step a dimer is introduced in the junction of the $L$ and $R$ blocks, i.e., the $J$ block. The updated state turns out to be
\begin{equation}
|N-2,0\rangle_{L}~ |2,0\rangle_{R}\xrightarrow{k=1} |N-1,0\rangle ~|\chi_2\rangle ~|1,0\rangle,
\end{equation}
\end{enumerate}
%\begin{figure}
%
%  \includegraphics[angle=0,width=3.2cm]{fig2a.png}
%  \includegraphics[angle=0,width=4.5cm]{fig2b.png}
%     \includegraphics[angle=0,width=3.8cm]{fig3a.png}
%  \includegraphics[angle=0,width=4.cm]{fig3b.png}
% \caption{Schematic diagram of the junction terms a) $|\chi_2\rangle$ and b) $|\chi_3\rangle$ and periodic terms  c)~$|\xi\rangle$ and d)~$|\gamma_1\rangle, |\gamma_2\rangle$}
%  
%     \label{fig2}
%   \end{figure}
where the state $|\chi_2\rangle$ is depicted in Fig.~\ref{fig3h}(a). Now combining the above three steps, the final state after introduction of a single dimer in the lattice is given by
\begin{eqnarray}
|N,0\rangle &\xrightarrow{k=1}& |N-1,1\rangle\nonumber \\
&\equiv& |N-3,1\rangle _{L}~|2,0\rangle_{R} + |N-2,0\rangle_{L}~|1,1\rangle_R \nonumber\\
&+& |N-1,0\rangle~ |\chi_2\rangle~ |1,1\rangle.
\end{eqnarray}
                     
For example, consider the initial state $|4,0\rangle$ in Eq.~(\ref{example}). We have,
$|4,0\rangle=|2,0\rangle_L~|2,0\rangle_R$. Then the state, after introduction of one dimer, would be (see Fig.~\ref{fig4h} for an illustration of the three update paths)
\begin{eqnarray}\label{eqn:examp}
|4,0\rangle &\xrightarrow{k=1}& |3,1\rangle = |1,1\rangle~|2,0\rangle + |2,0\rangle~|1,1\rangle\nonumber \\
&+&|1,0\rangle~ |\chi_2\rangle~ |1,0\rangle,
\end{eqnarray}\\
where the first two terms are the contributions from the blocks $L$ and $R$, and the third term comes from the update of the junction, $J$.
 Now after completion of the first step, we need to introduce one more dimer into the present configuration in order to continue the iteration process. It can be done following a path similar to the one described above, i.e., a direct update of the $L$ and $R$ blocks, which is basically updating all the terms of the state by introducing dimers into the left and right blocks, and an update, which consists of injecting a dimer at the junction block. The above scheme can be repeated $k$ times so that the final state contains $k$ dimers and $2(N-k)$ holes in the lattice.
In general, by updating the $L$, $J$, and $R$ blocks, with $k'$ = 1 singlets, we obtain the recursive generator expressed in Eq.~(\ref{eq:generator}).
%to obtain the RVB state, which can be written as
%\begin{eqnarray}
%|N-k,k\rangle &=& \mathcal{U}^{\otimes{k'=1}}|N-k+1,k-1\rangle + |N-k-1,0\rangle\nonumber\\
%&\times&  |\chi_{k+1}\rangle
%+ |N-k-2,0\rangle |\chi_{k+1} \rangle |1,0\rangle,
%\label{eqn:recur_state}
%\end{eqnarray}
%
%%% 
%%\begin{equation}
%% \begin{split}\label{eqn:recur_state}
%% |(N-k)',(k)\rangle= \oplus|(N-k+1)',(k-1) \rangle+\\
%% |(N-k-1)'\rangle |\chi_{k+1}\rangle +|(N-k-2)'\rangle |\chi_{k+1} \rangle |(1)',0\rangle,
%% \end{split}
%% \end{equation}
As mentioned before, here $\mathcal{U}^{\otimes{k'}}$ is the direct update operator to inject $k'$ dimers in the $L$ and $R$ blocks of the state, $|N-k+1,k-1\rangle$.
Subsequently, the second and the third terms in Eq.~(\ref{eq:generator}) correspond to the indirect update of the $J$ block. For example, the first two terms in Eq.~(\ref{eqn:examp}), $|1,1\rangle~|2,0\rangle$ and $|2,0\rangle~ |1,1\rangle$, is generated from the direct update of the state $ |4,0\rangle$ and the third term $|1,0\rangle~ |\chi_2\rangle |1,0\rangle$ emerges from the indirect update of the junction sites. Note that, there may arise similar terms due to the update process of the $L$, $R$, and the $J$ block. In those cases we need to carefully include such terms  only once in the recursion, so that overcounting of the terms can be avoided. In Eq.~(\ref{eq:generator}), we note that the term $|\chi_{k+1}\rangle$ can be generated recursively from $|\chi_{k}\rangle$ by introducing an additional rung to the left and assigning a dimer along the horizontal direction, as demonstrated in Fig.~\ref{fig3h}, for $|\chi_{2}\rangle$, $|\chi_{3}\rangle$ and $|\chi_{4}\rangle$.

 In the succeeding section, we present a detailed discussion on how reduced density matrices for a block of lattice sites can be obtained from the
recursion method. For the purposes of our study, a block of four sites, in two nearest neighbor (NN) rungs of  the ladder, is sufficient. 

%\vspace{-.2cm}
\section{ Recursion relation for reduced density matrices}
\label{recursion_reduced}
In order to calculate the $\mathcal{G}$ of the doped RVB state, we required to derive expressions for the reduced density matrices, using the generator expressed in Eq.~(\ref{eq:generator}).
%{\color {red} However, as mentioned before, even for a moderate-sized systems, it becomes extremely difficult to consider all possible dimer coverings and formulate the analytical form of the reduced density matrix from there. In order to overcome such computational complexities, here we propose an analytical recursion relation for the reduced density matrix with small number of lattice sites,   obtained from a large doped  ladder.}
Let us consider the cases for open and periodic ladders separately.

\noindent a) {\it Open ladder}: The primary method to build the recursive relations is to divide the lattice into blocks and junctions. The advantage lies in the fact that these blocks do not overlap, and hence can be independently traced to obtain ${\rho_{red}}$ needed to calculate the $\cal{G}$. Hence, from the non-periodic ladder state, $|N-k,k\rangle$, by tracing all sites apart from rungs $m-1$ and $m$, we get the reduced state, $\rho_{red}^{\cal{NP}}$, of 4-sites , given by
%\begin{figure*}[h!]
%\vspace{-1cm}
\begin{widetext}
%\vspace{-1cm}
\begin{eqnarray}
%\label{red_np}
%\label{red_np}
\boldsymbol{\rho^{{\cal{N}{P}}}_{red}} &=&  \sum^2_{i=0} \mathcal{Z}^{(\mathcal{S}-1+i)}_{(k-i)} |2-i,i\rangle \langle 2-i,i| +%\nonumber\\
%&+&
\sum^{k+1}_{k_1=2} \mathcal{Z}^{(\mathcal{S}-1)}_{(k-k_1+1)}~\text{tr~}(|\chi_{k_1} \rangle \langle \chi_{k_1}|)|1'\rangle \langle 1'|+
%\nonumber\\
%&+&
\sum^{k}_{k_2=2} \mathcal{Z}^{(\mathcal{S})}_{(k-k_2)}~\text{tr~}(|\chi_{k_2} \rangle \langle \chi_{k_2}|) |1\rangle \langle 1|\nonumber\\
&+&\sum^1_{i=0} \mathcal{Z}^{\substack{(\mathcal{S}+i)}}_{(k-2-i)} \text{tr~}(|\bar{2}\rangle \langle \bar{2}|) |1-i,i\rangle \langle 1-i,i|
%\nonumber \\
%&+&
+\sum^{k+1}_{k_3=3} \mathcal{Z}^{(\mathcal{S})}_{(k-k_3+1)}~ (\text{tr~}(|\chi_{k_3} \rangle \langle \chi_{k_3}|))
%\nonumber\\
%&+&
+\sum^{k}_{k_4=2} \mathcal{Z}^{\substack{(\mathcal{S})}}_{(k-k_4)}(\text{tr~}(|\chi_{{k_4}+1}\rangle \langle \chi_{k_4}| \langle 1|) \nonumber \\
&+& \text{h.c.})+\sum^{\substack{i=1,\\j=k-2-i}}_{i=0,j=0} \mathcal{Z}^{(\mathcal{S}+i)}_{(k-2-i-j)}~(1/2)^{j+1}
%\nonumber\\
%&\times&
(|1\rangle |1-j,j\rangle \langle 1-j,j+1|+\text{h.c.}), \textrm{where $\mathcal{S}$ = $N-k-1$, and}
\label{eq1}
%%%%NEW%%%%
\end{eqnarray}
%\end{figure*}
%\vspace{-3cm}
\end{widetext}
%&+& \frac{1}{4} \sum_{i=0,j=0}^{i=1,j=k-3} \mathcal{Z}^{(\mathcal{S}+i)}_{(k-4-i-j)}
%\nonumber \\
%&\times&
%\left(\frac{1}{2}\right)^j  \text{tr}~|1\rangle |1-i,i\rangle |1\rangle \langle \bar{2}|\langle 1-i,i|
%\nonumber \\
%&+&
%\frac{1}{4}    \sum^{\substack{j=k-5,\\i=k-5+j}}_{j=0,i=0} \mathcal{Z}_{k-5-(i+j)} \left(\frac{1}{2}\right)^{i+j+1} |1\rangle |1'\rangle |1\rangle \langle\bar{2}| \langle 1'|, \label{eq2}\\\nonumber\\
%%%%%NEW
%\hspace{3cm}
$\mathcal{Z}^{N-k}_k=\langle N-k,k|N-k,k\rangle$, and $|\bar{2}\rangle$ = $|0,2\rangle - |0,1\rangle |1,0\rangle$. Numerical studies for a moderate $N$, suggest that obtaining the reduced state of a square block of 4 sites for large ladders, which is symmetric for the ladder, is sufficient for the computation of $\mathcal{G}$. Hence, we use the recursion method to obtain the 4-site reduced state (${\rho_{red}}$) at rungs $m-1$ and $m$.

The main advantage in formulating the  recursion relation for the entire state, as expressed in Eq.~(\ref{eq:generator}), can be seen when one needs to obtain the reduced density matrix, $\rho_{red}$. This is because the terms which correspond to the blocks $R$ and $L$ are mutually orthogonal to those belong to the junction block $J$. As a result, in the expression for $\rho_{red}$, one would never get any contribution from the terms that emanate from $|{\raisebox{-1ex}{\scalebox{3}{$\cdot$}}}\rangle_{L(R)}\langle {\raisebox{-1ex}{\scalebox{3}{$\cdot$}}} |_{J}$. \
Now if one starts from the $L$ and $R$ blocks coverings and traces out all but the sites of last two rungs (sites (1-4) in Fig.~\ref{fig2h}), then there would be the following three possibilities,
 
~~~i) The reduced block contains holes only.
 ~~ii) The reduced block contains one singlet and one pair of hole.
 ~~iii) The reduced block contains singlet coverings only.\\
 \noindent The first term in Eq.~(\ref{eq1}),
\begin{equation}
\sum^2_{i=0} \mathcal{Z}^{(\mathcal{S}-1+i)}_{(k-i)} |(2-i),(i)\rangle \langle (2-i),(i)|\nonumber
\end{equation}
basically corresponds to the above possibilities. As an example, consider an initial eight-site doped RVB state which includes only one singlet and 3 hole-pairs. The contribution from
the $L$ block and $R$ block would lead to the following terms in the expression of the doped RVB state,
$ |1,1\rangle_{L}\otimes |2,0\rangle_{R} + |2,0\rangle_{L}\otimes|1,1\rangle_{R}$. %+|1',0\rangle_{L}\otimes|\chi_2\rangle_J\otimes|1',0\rangle$. Hence the
Therefore, the reduced state would contain following terms, $a_1~ |2,0\rangle\langle2,0|+a_2~|1,1\rangle\langle1,1|$,
%\begin{eqnarray}
%\rho^{\mathcal{N}\mathcal{P}}_{red}&=&a_1~ |2,0\rangle\langle2,0|+a_2~|%1,1\rangle\langle1,1|,\nonumber
%     && \nonumber,%+a_3~\text{tr}(|\chi_2\rangle\langle\chi_2|)|1',0\rangle \langle 1',0|%\nonumber,
%\end{eqnarray}
where $a_1$ = $\mathcal{Z}^{1}_{1}$, and $a_2$ = $\mathcal{Z}^{2}_0$, which can be  obtained from Eq.~({\ref{eq1}).

Subsequently, the junction $J$ would generate  additional terms in the expression of the reduced state. As an example, first consider terms which has only one singlet at the junction block (see Fig.~\ref{fig3h}). Mathematically, those can be expressed as $|{\raisebox{-1ex}{\scalebox{3}{$\cdot$}}}\rangle \otimes |\chi_k\rangle \otimes
|{\raisebox{-1ex}{\scalebox{3}{$\cdot$}}}\rangle$, where $|\chi_2\rangle$ and $|\chi_3\rangle$ are depicted in Fig.~\ref{fig3h}. Now the contributions from the overlap of those terms are given by second, third, and fourth terms in Eq.~(\ref{eq1}). Considering, once more,  the previous example of an eight-site doped RVB state containing only one singlet, we can write the contributing term
from the junction as $|1,0\rangle\otimes |\chi_2\rangle\otimes|1,0\rangle$. Hence after tracing out all but the sites those are at the last two-rungs, we get
\begin{eqnarray}
\rho^{\mathcal{N}\mathcal{P}}_{red}=a_1~ |2,0\rangle\langle2,0|+a_2~|1,1\rangle\langle1,1|\nonumber \\+ a_3~ \text{tr}~|\chi_2\rangle\langle\chi_2|\otimes |1,0\rangle \langle 1,0|,
\end{eqnarray}
where $a_3$ = $\mathcal{Z}^{1}_{0}$ is again evaluated using Eq.~(\ref{eq1}).

\begin{figure}[h]
\begin{center}
\includegraphics[angle=0,width=8.5cm]{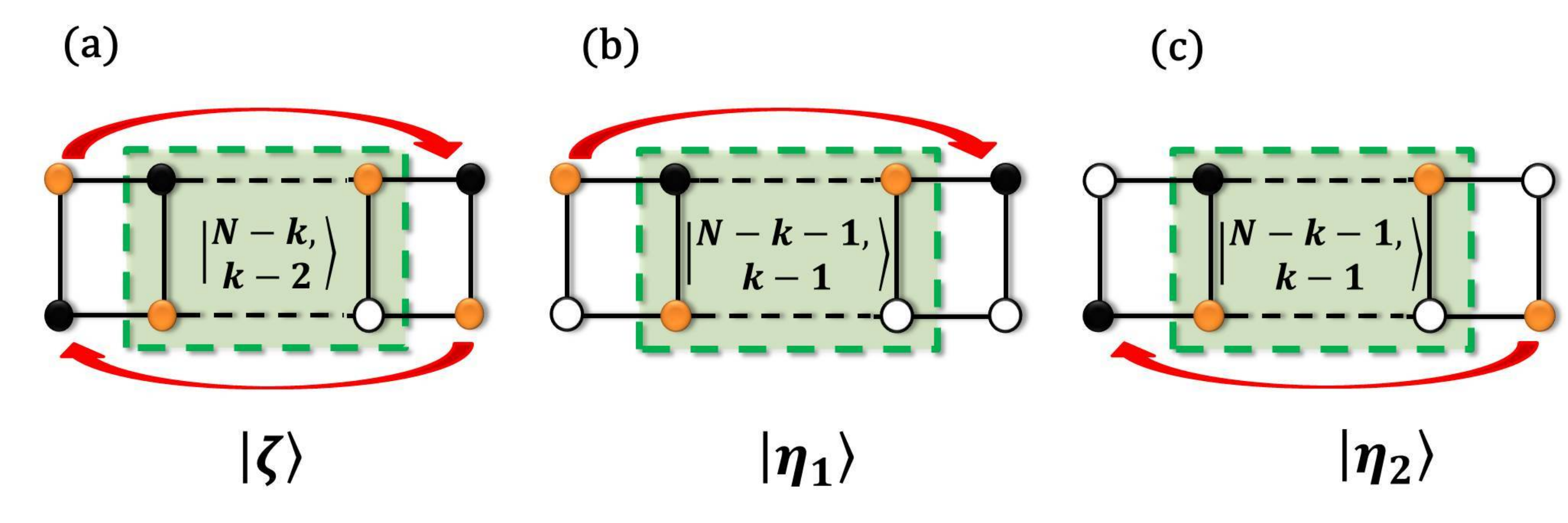}
\caption{Schematic diagram of the blocks $L$, $R$, and $J$ in the spin lattice with periodic boundary condition.}
\label{fig5h}
\end{center}
%\vspace{-.2cm}
\end{figure}

Additionally, there may be terms which would contain two horizontal singlets at the junctions such as $|{\raisebox{-1ex}{\scalebox{3}{$\cdot$}}}\rangle \otimes|\bar{2}\rangle \otimes|{\raisebox{-1ex}{\scalebox{3}{$\cdot$}}}\rangle$. Those would
certainly have non-zero overlap with the $L$ and $R$ block terms. The fifth and sixth terms of Eq.~(\ref{eq1})  correspond to contribution from these terms. As an example,
instead of inserting one singlet if we now introduce two singlets in the eight-site doped RVB state i.e. $|2,2\rangle$, we would have terms in the doped RVB state such as
$|1,0\rangle \otimes |\bar{2}\rangle\otimes |1,0\rangle$. Hence the reduced density matrix would have following terms,
\begin{eqnarray}
a_4~\text{tr}~|\bar{2}\rangle\langle\bar{2}|\otimes |1,0\rangle \langle 1,0|+a_5/2~|0,1\rangle|1,0\rangle \langle H|, \nonumber
\end{eqnarray}
where $a_4$ and $a_5$ are given by $\mathcal{Z}^{1}_{0}$ which can be obtained from Eq.~(\ref{eq1}), for $N=4$ and $k=2$.
%$a_3=\mathcal{Z}^{1'}_1$.\\

%Moreover, $|2\rangle$ is the undoped RVB state for a 2-runged ladder comprised with two singlets and the states $|1\rangle$ and $|1'\rangle$ are the singlet and  hole-pair along the legs of the ladders  respectively.
\noindent b) {\it Periodic ladder}: Incorporation of periodic boundary condition, $N$ + $1$ $\equiv$ $N$, leads to additional terms in the state $|N-k,k\rangle$.
Consequently, we find that the number of terms in the expression of the recursively generated reduced density matrix, $\rho^{\cal{P}}_{red}$, increases, due to overlap of different states. These extra terms in $\rho^{\cal{P}}_{red}$ can be redeemed by analyzing two separate situations, say ${\cal{P}}_1$ and ${\cal{P}}_2$, corresponding to different boundary terms:\\ %as described in the supplementary material.

\noindent I) $\mathcal{P}_1$ -- the end sites of both the legs share a singlet (see Fig.~\ref{fig5h}(a)). \\

\noindent II) $\mathcal{P}_2$-- the end sites of only one of the leg share a singlet and the end
site of the other leg contains holes (see Figs.~\ref{fig5h}(b) and (c))). \\

Therefore, in the expression of the reduced density matrix, there would be additional terms due to the overlap of the
states in I with itself, and the $L$ and $R$ blocks ($\rho^{\mathcal{P}_{1}}_{red}$) (see Eq.~(\ref{eq2})) and similarly, the overlap of states in II with itself, and $L$ and $R$ blocks $(\rho^{{\mathcal{P}}_2}_{red})$ (see Eq.~(\ref{eq3}) for $N-k=1$ and Eq.~(\ref{eq4}) for $N-k=2$).
%\vspace{-.1cm}
\begin{widetext}
\begin{eqnarray}
\boldsymbol{\rho^{{\cal{P}}_1 }_{red}} &=& \text{tr}~ (|\zeta\rangle \langle \zeta|)+
\frac{1}{2}  \sum_{i=0,j=0}^{\substack{i=1,\\j=k-2-i}} \mathcal{Z}^{\substack{(\mathcal{S}+i)}}_{(k-2-i-j)}\left(\frac{1}{2}\right)^j |1-i,i\rangle |1\rangle\langle i-1,i+1| +
\frac{1}{2}  \sum_{j=0}^{j=k-4} \mathcal{Z}^{\substack{(\mathcal{S}+1)}}_{(k-4-j)} \left(\frac{1}{2}\right)^j
\text{tr}~(|\bar{2}\rangle \langle \bar{2}|)~|1\rangle \langle 1| \nonumber\\
%\nonumber\\
&+&
\left(\frac{1}{2}\right)^j  \sum_{k_6=2,j=0}^{\substack{k_6=k-1,\\j=k-k_6-1}} \mathcal{Z}^{\substack{(\mathcal{S})}}_{(k-k_6-1-j)}
\left(\frac{1}{2}\right)^j  \text{tr}~(|\chi_{k_6}\rangle \langle \chi_{k_6}||1\rangle \langle 1|+ |\chi_{k_6}\rangle |1\rangle \langle \chi_{k_6+1}|)
%%%%\nonumber\\
%&+&
%\frac{1}{2}  \sum_{k_1=2,j=0}^{\substack{k_1=k-1,\\j=k-k_1-1}} \mathcal{Z}^{\substack{(\mathcal{S})}}_{(k-k_1-1-j)}
%\left(\frac{1}{2}\right)^j \text{tr}|\chi_{k_1}\rangle |1\rangle \langle \chi_{k_1+1}|
%&+& \frac{1}{2}   \sum_{j=0}^{j=k-4} Z_{(N-k)',(k-4-j)} \left(\frac{1}{2}\right)^j \text{tr}~|\bar{2}\rangle|1\rangle \langle 1| \langle 2|\nonumber\\
%%%%&+&
+\frac{1}{2}   \sum_{i=0,j=0}^{j=k-4-i} \mathcal{Z}^{(\mathcal{S}+1)}_{(k-4-(i+j))} \left(\frac{1}{2}\right)^{i+j}
\nonumber \\ &\times&
\text{tr}~(|\bar{2}\rangle|1\rangle \langle 1| \langle 2|)+\frac{1}{2}  \sum_{i=0,j=0,l=0}^{\substack{i=1\\j=k-3-i\\l=k-3-(i+j)}} \mathcal{Z}^{\substack{(\mathcal{S}+i)}}_{(k-3-(i+j+l)} \left(\frac{1}{2}\right)^{j+l}~\text{tr}~(|1\rangle |1-i,i\rangle |1\rangle \langle \bar{2}|
%\nonumber \\&\times&
\langle 1-i,i|) \label{eq2}, \textrm{where $\mathcal{S}$ = $N-k-1$,}\\
\boldsymbol{\rho^{{\cal{P}}_2}_{red}} &=& |\gamma_1\rangle \langle \gamma_1|+|\gamma_2 \rangle \langle \gamma_2|+\left(\frac{1}{2}\right)^{k-3}( |\gamma_1\rangle \langle \gamma_2|+|\gamma_2 \rangle \langle \gamma_1| )
%\nonumber\\&&
%\hspace{2cm}
+ {\bf{\big[}}(|\gamma_1 \rangle +|\gamma_2\rangle) (\mathcal{D}_{k-2}  \langle 1| \langle \bar{2} | \langle 1'|+\mathcal{D}_{k-1}  \langle H|)+h.c.{\bf{\big]} } (-1)^{{k+1}/{2}},\nonumber\\
&&\textrm{where $N-k$ = 1,}\label{eq3} \\%\forall ~~N-k = 1
%%%%%NEW
%\hspace{3cm}
\boldsymbol{\rho^{{\cal{P}}_2}_{red}} &=& (|\xi \rangle \langle \xi|+|\xi\rangle \langle \chi_2| \langle 1'|) (-1)^{k/2 -1} \mathcal{D}_{k-1},~~
\textrm{where $N-k$ = 2.}
\label{eq4}%~\forall ~~N-k = 2
\end{eqnarray}
%\vspace{-0.2cm}
%\end{figure*}
%\vspace{-.2cm}
\end{widetext}

%Moreover, some additional terms may appear due to the overlap between the blocks I--II $(\rho^{{\mathcal{P}}_{12}}_{red})$. These terms will accordingly contribute to the final reduced state, $\rho_{red}$.
The terms $|\xi\rangle$, $|\gamma_{1}\rangle$, $|\gamma_{2}\rangle$, and $|\zeta\rangle$, along with an illustrative description of the recursion method is provided in the Figs.~\ref{fig3h} and~\ref{fig5h}.  ${\mathcal{D}_k}$ can recursively be generated using ${\mathcal{D}_x}$ = ${\mathcal{D}_{x-1}}+2{\mathcal{D}_{x-2}}$ with the initial condition ${\mathcal{D}_0}$ = ${\mathcal{D}_1}$ = $1$.
Now if $N-k=1$, periodic states corresponding to the two types, ${\cal{P}}_1$ and ${\cal{P}}_2$, would overlap with each other and lead to the following additional terms in the expression of the total reduced density matrix of 4-sites given by,
$\rho^{{\cal{P}}_{12}}_{red}$ = $1/2^{(k-3)} (-1)^{(k+1)/2}( |\gamma \rangle ( \langle \gamma_1| $ + $ \langle \gamma_2 |)+ h.c.)  \mathcal{D}_{k-3}$,
where $|\gamma\rangle$ = $|\bar{2}\rangle_{1, N}|N-k-2,k-2\rangle_{2, N-1}$.
Hence considering all possible periodic boundary terms, the expression of the reduced density matrix for the system is given by
$ \rho^{{\cal{P}}}_{red}$ = $\rho^{{\cal{N}{P}}}_{red}$ + $\rho^{{\cal{P}}_{1}}_{red}$ + $\rho^{{\cal{P}}_{2}}_{red}+\rho^{{\cal{P}}_{12}}_{red}$.
Fig.~\ref{fig:ggm} shows the $\cal{G}$ of a periodic doped RVB state calculated using the recursion method for upto $300$ lattice sites.

 The expression of the reduced density matrices obtained using the recursion method can be applied to compute various bipartite as well as multipartite physical quantities that characterize the ground state properties of the system, even for large lattice size. In the following section, we will look at the genuine multipartite entanglement properties of the
doped RVB ladder, which can be efficiently obtained using this technique.

%%  \begin{equation}\label{rho_final}
%%  \rho^{{\cal{P}}}_{red}=\rho^{{\cal{N}{P}}}_{red}+\rho^{{\cal{P}}_{1}}_{red}+\rho^{{\cal{P}}_{2}}_{red}+\rho^{{\cal{P}}_{12}}_{red}.
%%  \end{equation}

%{\it Recursion relation for reduced density matrices}:

\section{Genuine multiparty entanglement in quantum ladders}
\label{genuine}
Here we investigate the multipartite entanglement of a doped quantum spin ladder, under the RVB ansatz. Since the study of GS properties of $t$-$J$ Hamiltonian is limited to numerical simulations and approximate methods, explicit estimation of multipartite entanglement is extremely difficult for large systems. The doped RVB ansatz for the GS of the $t$-$J$ model provides a viable alternative to study such quantities.
%Moreover, our aim is to check whether the RVB GS can effectively capture the trends of multipartite entanglement for the $t$-$J$ model.
%We again use the qutrit basis for a joint description of the dimer-hole system.
It is known that the RVB liquid state with no holes, $|\Psi\rangle_{N,0}$, is rotationally invariant under the unitary $U^{\otimes 2N}$, where $U$ is a local unitary acting on a single qubit~\cite{ours3,chan,ours1}. In the composite dimer-hole qutrit space, the doped RVB state, $|\Psi\rangle_{k,N-k}$, is invariant under unitary operations of the form $\tilde{U}^{\otimes 2N} = (U \oplus~ \mathbb{I})^{\otimes 2N}$, where $\oplus$ is the direct sum, $\mathbb{I}$ is the scalar 1 and $U$ is an arbitrary single qubit unitary. This invariance property of doped RVB ladders is important in investigating its multipartite entanglement as shown below.

{\bf Theorem:} The doped RVB ladder state, $|\Psi\rangle_{k,N-k}$, with $2N$ lattice sites, containing all possible coverings of $k$ ($k \neq$ 0) spin dimers
interspersed with $2(N-k)$ holes, is always genuinely multipartite entangled
for all ladder topologies that are periodic or infinite along the rails.

{\bf Proof:} To prove that $|\Psi\rangle_{k,N-k}$ is genuinely multisite entangled, we need to show that the state is entangled across every possible bipartition or alternatively, we have to prove that all reduced density matrices of the system are mixed.
Using the invariance of $|\Psi\rangle_{k,N-k}$ under the action of $\tilde{U}^{\otimes 2N}$, one can show that
all $p$-qutrit reduced systems, $\rho^{(p)} = \textrm{Tr}_{\bar{p}}[|\Psi\rangle\langle\Psi|_{k,N-k}]$, obtained by tracing over all but $p$ ($\bar{p}$) sites, is always invariant under $\tilde{U}^{\otimes p}$.
%Using the above lemma, one can show that
Hence, a single qutrit reduced state
%, invariant under $\tilde{U}^{\otimes p}$,
must have the form,
$\rho^{(1)}$ = $p|2\rangle\langle 2|$ + $(1-p)/2~\mathbb{I}_2$, where $\mathbb{I}_2$ = $|0\rangle\langle 0|$ + $|1\rangle\langle 1|$ and $p$ is fixed by the number of holes in the system. The relation shows that $\rho^{(1)}$ is always mixed for $p \neq$ 1.
Since
%the periodicity and symmetry of the doped RVB ladder ensures that
all $\rho^{(1)}$ are equivalent, the condition $p$ = 1 is satisfied \textit{iff} all $2N$ sites contain holes.
%, and $|\Psi\rangle$ is a trivial multiparty separable state.
Similarly, the nearest neighbor two-site density matrix  has the form,
$\rho^{(2)}$ = $p_1|2 2\rangle\langle 2 2| + p_2/9~\mathbb{I}_9 + p_3 W_2(q)$, where $\mathbb{I}_9$ is the identity matrix on $\mathbb{C}^3 \otimes \mathbb{C}^3$ and $W_2(q)=q|\psi^-\rangle \langle \psi^-|+(1-q)\mathbb{I}_4/4$ is the Werner state \cite{wer} with $\mathbb{I}_4$ being the identity operator on the 4-dimensional space  defined in the projected two-qubit spin basis.
% with parameter $q$.
%Hence, both the single and two qutrit reduced state are mixed for a finite electron density.
Now, $\rho^{(2)}$ is pure when $p_1=p_2=0$ and $q=1$. Which implies that it is pure \textit{iff} the entire lattice is either  filled with holes or is a single dimer covering, and these options are disallowed by the premise. Therefore, $\rho^{(1)}$ and $\rho^{(2)}$ are always mixed and $|\Psi\rangle_{k,N-k}$ is always entangled across these bipartitions.

However, we want to show that all possible bipartitions, irrespective of the number of sites, are always mixed. To prove this let us
assume now that an arbitrary $p$-site density matrix ($\rho^{(p)}$) is pure, which implies that $|\Psi\rangle_{k,N-k}$ is separable along that $p$: $2N-p$.
Let $p$ = $p_1$ + $j$, where $j$=$1$ or 2 such that $|j|<|p_1|$ ($|\cdot|$ is the cardinality of the argument).
For the periodic or infinite ladder, %is isotropic along the vertical and periodic along the horizontal,
one can always find another equivalent pure density matrix, $\rho^{(q)}$, such that $q$ = $q_1$ + $j$ and $|p|=|q|$, where $j$-sites overlap. By assumption, both $\rho^{(p)}$ and $\rho^{(q)}$ are pure. Using strong subadditivity of von Neumann entropy, $S(\sigma)=-\text{tr}(\sigma~ \text{log}_2 \sigma)$~\cite{ss}, we obtain
$
S(\rho^{(p_1)})+S(\rho^{(q_1)}) \leq S(\rho^{(p_1 + j)})+ S(\rho^{(q_1 + j)}).
%\label{VN}
$
%where \(S(\cdot)\) denotes the von Neumann  entropy.
Now $S(\rho^{(p_1+j)})$ = $S(\rho^{(q_1+j)})$ = 0, since $\rho^{(p)}$ and $\rho^{(q)}$ are pure. Since $S \geq$ 0, we have  $S(\rho^{(p_1)})$ = $S(\rho^{(q_1)})$ = 0, and therefore $S(\rho^{(j)})$ = 0 implying $\rho^{(j)}$ is pure, which is not true since all $\rho^{(1)}$ and $\rho^{(2)}$ are mixed under finite doping. The contradiction implies that all reduced density matrices, $\rho^{(p)}$ are mixed and all $p$: $2N-p$ are entangled.

We note that the above proof does not include the $p$: $2N-p$ bipartitions where no equivalent $\rho^{(q)}$ with overlap is feasible, such as the bipartition between the two legs of the ladder. However, in such cases, the theorem can be proved using a different argument. We assume that the legs, $L_i$ and $L'_i$ of $|\Psi\rangle_{k,N-k}$ are pure and thus the entire state is separable along that $N$: $N$.
%So, $\rho^{(N)}_{(L_1:L_N)}$ and $\rho^{(N)}_{(L'_1:L'_N)}$ are pure.
For the above condition to be satisfied, all reduced states along the rungs, $\rho^{(2)}_{(L_k,L'_k)}$, $\forall~ k$,  must be separable. However, as can be shown by using recursive method, such nearest-neighbor $\rho^{(2)}$ states are always entangled. Hence, the doped RVB state is genuinely multipartite entangled.
\hfill\(\blacksquare\)

%
%%\section{Generalized geometric measure}
%%\label{ggm_def}
% In the previous section, we observe that the doped RVB state formed using Eq.~(\ref{Eqn:RVB}), is indeed genuinely multiparty entangled.
 Let us now quantify the genuine multipartite entanglement in  doped RVB ladders and 
%At this stage, the very next thing one should look for is to
characterize
%how the multiparty entanglement is actually
its variation with the electron density.
Towards that aim, one needs to find a computable measure of genuine multiparty entanglement, which in our work
%is}  The genuine multisite entanglement measure considered in our work
is the ``generalized geometric measure'' (GGM)\cite{ggm1} (cf.\cite{ggm2}).  In  the forthcoming section, we provide all the necessary details required to compute GGM for any arbitrary $N$-party pure quantum state.
\section{Geuine multisite entanglement measure}
\label{ggm_definition}
The GGM, $\mathcal{G}$, of an N -party pure quantum state $|\phi\rangle$ is a computable measure and is basically the optimized fidelity distance of the state from the set of all states that are not genuinely multiparty entangled. In particular, the GGM $\mathcal{G}(|\phi\rangle)$ can be evaluated as
\begin{eqnarray}
\mathcal{G}(|\phi\rangle)=1-\lambda_{\text{max}}^2,
\end{eqnarray}
where $\lambda_{\text{max}}$ = max $|\langle\xi_n|\phi\rangle|$, $|\xi_n\rangle$ is an $N$-party non-genuinely multisite entangled quantum state and the maximization is performed over the set of all such states. For pure quantum states, it was shown that  GGM can be effectively computed using the straightforward relation~\cite{ggm1}
\begin{eqnarray}
\mathcal{G}(|\phi\rangle) &=& 1-\textrm{max}\{\lambda^2_{A:B}  |A\cup B = {A_1 , \dots, A_N }, \nonumber\\
&&~~~~~~~~~~~~ A\cap B=\phi\},
\end{eqnarray}
where $\lambda_{A:B}$ is the maximum Schmidt coefficient in all possible bipartion split of $A:B$ of the given state $|\phi\rangle$.\

Genuine multipartite entanglement is a well understood physical property in entanglement theory (see Ref.~\cite{horodecki} ), which essentially captures the presence of entanglement between every constituent of a many-body system. In contrast, measures such as entanglement entropy and entanglement of formation are essentially bipartite entanglement measure, which do not necessarily say anything about the global entanglement properties of a many-body state. Although entanglement entropy is important in studying cooperative phenomena such as area laws, it is not adequate to study the multipartite entanglement properties of many-body systems. Presence of multipartite entanglement may give rise to interesting cooperative properties that are not necessarily exhibited by restricting to bipartite entanglement.  An obvious advantage of using GGM, is that it can be efficiently calculated through the reduced density matrices of a many-body quantum state.  One should stress here that entanglement entropy captures the distribution of entanglement between two blocks of the system constituted of connected cluster of spins. In comparison, GGM allows us to characterize the entanglement between all possible partitions of the system into two, three, four, ... blocks, comprising of connected as well as disconnected group of spins,   which provides insight about  the cooperative properties of the ground state, beyond correlation decay or area laws~\cite{multi_ent}.

\section{TRENDS OF GENUINE MULTISITE
ENTANGLEMENT: GS OF t-J LADDER Vs DOPED RVB STATE}
\label{model}
%Considering the short-range doped RVB state as a framework to study the GS entanglement properties of the $t$-$J$ model,
%we begin with
%
%
We now consider a quantum spin-1/2 ladder model, consisting of an arbitrary numbers of holes and spin particles,
and consider the short-ranged RVB state as a framework to study its GS multiparty entanglement properties.
%where the corresponding interaction is
%described by the $t$-$J$ Hamiltonian.
%
The model
%provides one of the best known theoretical frameworks to investigate high-$T_c$ superconductivity and
can be derived using second order perturbation theory from the Hubbard model in the limit of large on-site interaction.\cite{t-J,comm-tJ,t-J2,t-Janti} The $t$-$J$ Hamiltonian on a ladder can be written as
%\vspace{-.1cm}
\begin{equation}
\mathcal{H}= -t\sum_{\langle i,j\rangle,\sigma} {\mathcal{P}_G}~ (c_{i\sigma}^{\dagger}c_{j\sigma}+\text{h.c.}) ~{\mathcal{P}_G}+J\sum_{\langle i,j\rangle}\vec{S_i}\cdot\vec{S_j},
\label{Ham_t_J}
%\vspace{-.3cm}
\end{equation}
where $c_{i\sigma}$ ($c_{i\sigma}^\dag$) is the fermionic annihilation (creation) operator of spin $\sigma$ ($= \{\uparrow, \downarrow\}$), and $\vec{S}_i$ is the triad of spin-1/2 operators, at site $i$. The Heisenberg exchange coupling ($J$) is isotropic along the rungs and legs while $t$ represents the transfer energy and the expression $\langle i,j\rangle$ denotes that the sum is taken over nearest neighbor (NN) sites. $\mathcal{P}_G$ is the Gutzwiller projector $\Pi_i (1-n_{i\uparrow}n_{i\downarrow})$ which enforces at most single occupancy at each lattice site. This ensures that the undoped state physically represents a Mott insulator. The $t$-$J$ model, under finite doping, exhibits a rich phase diagram~\cite{t-J_phase,t-J_refernce, dimer-hole1, dimer-hole2,ll}.
% which has been extensively studied for low-dimensional antiferromagnets (AFM) \cite{t-J_phase,t-J_refernce, dimer-hole1, dimer-hole2}.
%In particular, the $t$-$J$ model in 1D and ladder configurations possess exotic correlation properties that are characterized by the Luttinger liquid theory \cite{ll}, as confirmed using exact diagonalization calculations, and exhibits  a rich superconducting phase for a  specific range of values of $J/t$ and electron density, $n_{el}$ \cite{sc_3,sc_4,t-J_refernce, dimer-hole1, dimer-hole2}.
%In this work, our interest lies in investigating $t$-$J$ ladders in the region $J/t \gtrsim 0.5$ where the superconducting phase seems to appear at relatively high $n_{el}$ \cite{t-J_phase}.
Note that these models can potentially be realized in fermionic ultracold gases at high energy scales~\cite{energy-scale}.
For moderate sized $t$-$J$ ladders, at half-filling, the Hamiltonian in Eq.~(\ref{Ham_t_J}) can be exactly diagonalized, provided certain properties of the system are invoked. For example, the spin number Hamiltonian, $\hat{N} = \sum_{i} (|0\rangle\langle 0| + |1\rangle\langle 1|)_i$, and the total spin along the z-axis, $\hat{S}^z = \sum_{i} S_i^z$ commute with the Hamiltonian, $\cal{H}$. Hence, the Hamiltonian can be block-diagonalized in the $(\mathbb{C}^3)^{\otimes 2N}$ Hilbert space basis for different total spin $\hat{S}^z$ and electron density $n_{el}=\langle \hat{N} \rangle /2N$. For our case, we assume that the spins form an initial insulating phase with $\hat{S}^z$ = 0, and with $n_{el}$ varying from 0 to 1. Note that $n_{el}$ = 0 and 1, correspond to a completely vacant and occupied lattice, respectively. For $n_{el}$ = 1, the state is an insulating RVB spin liquid. The doping concentration is denoted by $x$ = $1-n_{el}$. In our work, we have developed a numerical algorithm~\cite{comm2}, based on the Lanczos method~\cite{lanc}, to exactly solve the composite hole-dimer qutrit system. By dividing the Hilbert space in different subspaces, according to the hole concentration $x$ and total $\hat{S}^z$, exact ground state of the $t$-$J$ Hamiltonian can be obtained for upto 14 qutrits, with even number of holes.
%The algorithm is implemented using codes written in MATLAB and Fortran90.

%\section{Trends of GGM in doped RVB state and GS of $t$-$J$ Hamiltonian}
%\label{comparision}
Although we have shown that the doped RVB state is always genuinely multiparty entangled, a quantitative analysis of $\cal{G}$ requires its computation for large systems. Using the analytical recursion method proposed in the work, one can  recursively build the doped RVB state and subsequently obtain its relevant reduced density matrices which is necessary to estimate $\cal{G}$.
Figure \ref{fig:ggm} shows the behavior of the $\cal{G}$ with increasing $n_{el}$.
We observe that at $n_{el}$ = 0, $\cal{G}$ vanishes as expected since it corresponds to a product state, containing only holes. The maximum $\cal{G}$ is achieved at a critical density, $n_{el}$ = $n_{c} \approx$ 0.56.
%We observe that the maximum $\cal{G}$ occurs at $n_{el}$ = $n_c \approx$ 0.56.
%Interestingly, we find that this critical value of $n_c$ corresponds to the superconducting phase of the $t$-$J$ model.
Interestingly, we find that this critical value of $n_c$ with respect to
${\mathcal G}$ corresponds to that of the superconducting phase of the $t$-$J$
model.
% as discussed below.}

\begin{figure}[h]
\includegraphics[angle=0,width=6cm]{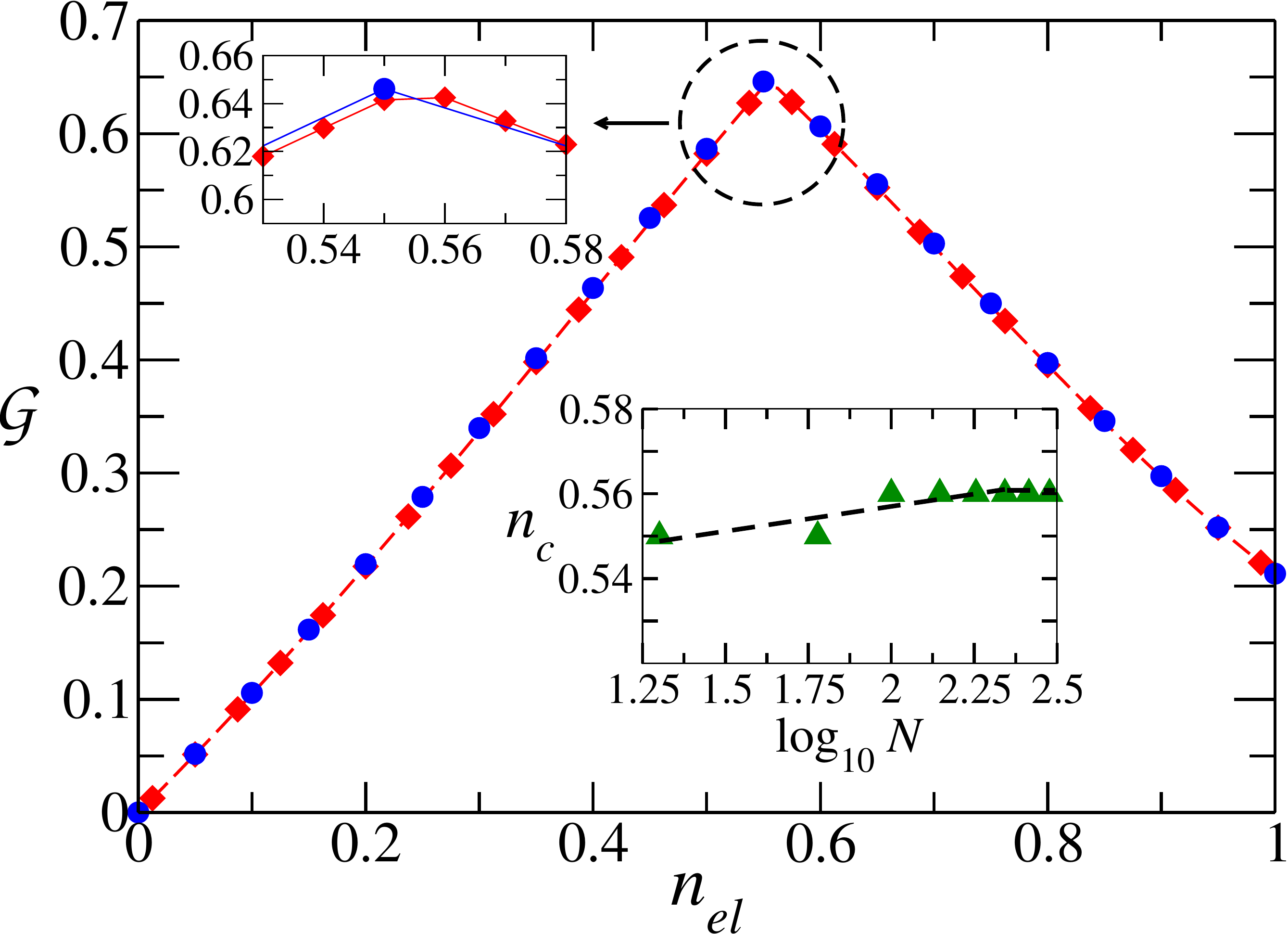}
\caption{(Color online.) Genuine multisite entanglement in doped RVB ladder. Variation of ${\mathcal{G}}$ with $n_{el}$ in doped RVB ladder states, for $2N$ = 40 (blue circles) and $200$ (red diamonds) lattice sites. The top inset magnifies the encircled region in the plot. The bottom inset shows the scaling of $n_c$ with the $\log_{10}$~$N$. The inset shows that as $N$ increases (plotted up to 300 sites), $n_c$ converges to 0.56. This is to be compared to the result for the systems described by the $t$-$J$ model in the superconducting regime, in Fig.~\ref{fig1}. }
\label{fig:ggm}
%\vspace{-.5cm}
\end{figure}
% Followingly, a recursion relation for the reduced density matrix comprising of small number of lattice sites is also obtained which indeed
% can be useful in order to find different observables characterizing the system.

%The GS of the $t$-$J$ Hamiltonian calculated using exact diagonalization algorithm  may lead to degenerate ground state for certain parameters values. To overcome the degeneracy, a finite interaction is introduced in the $z$-direction of the Heisenberg part of the Hamiltonian, making it an \textit{XXZ} model. We then estimate $\cal{G}$ for upto 14 qutrits.

We now consider the behavior of genuine multisite entanglement in the GS of the periodic $t$-$J$ ladder, obtained through exact diagonalization. Figure \ref{fig1}, shows the variation of $\mathcal{G}$ with $n_{el}$ for different moderately-sized systems. We observe that the behavior of $\mathcal{G}$ is qualitatively equivalent to those corresponding to doped RVB states  %\textcolor{red}{
(see Fig.~\ref{fig:ggm}). Below a certain critical density ($n_{c}$), i.e., in the region where $n_{el}< n_c$, $\cal{G}$ scales linearly with $n_{el}$, independent of $J/t$. This is due to the fact that $\cal{G}$ is obtained from  the $1$:rest bipartition, where the single-site density matrix is diagonal in the  computational  basis, with elements $[1-n_{el}, n_{el}/2, n_{el}/2]$. When $1 \geq n_{el} \geq n_c$, the $\cal{G}$ is a function of both $n_{el}$ and $J/t$. For the $t$-$J$ ladders, the maximum $\cal{G}$ is achieved at a critical density, $n_{c} \approx$ 0.65 which is close to that obtained using the doped RVB state ansatz. However, the small discrepancy in the exact values of the  electron densities  need to  account for finite size effect.
%}Even though our analysis with $t$-$J$ ladders is restricted to moderate system size, it is evident from Figs.~\ref{fig:ggm} and \ref{fig1} that doped RVB ansatz effectively encapsulates multipartite features of the exact GS of the $t$-$J$ ladders.
%\textcolor{red}{
Moreover, we infer that this critical density mark the onset of a superconducting phase in the two-legged $t$-$J$ ladder. For example in $t$-$J$ ladder with $J/t\approx 0.6$, the superconducting phase has been predicted to occur for relatively high values of $n_{el}$~\cite{t-J_phase}, which is close to the critical density corresponding to the $\cal{G}$, as obtained from our analyses.  

Even though the microscopic theory behind high-$T_c$ superconductivity~\cite{sc_1,sc_2,sc_3,sc_4,sc_5} remains unresolved~\cite{knowno,Mott}, $t$-$J$ ladder stands out as an important framework for understanding this novel phenomena~\cite{Anderson_tJ,tj}. Furthermore, the short-range RVB ansatz has been pitched to describe the superconducting states of the $t$-$J$ ladder~\cite{anders_rvb, Anderson_RVB}. The RVB state is a possible GS of the half-filled $t$-$J$ ladder~\cite{white} and, upon finite doping, provides a simple mechanism to describe high-$T_c$ superconductivity.  In this respect, our work indicates that $\cal{G}$ bears the signature of the $t$-$J$ ladder entering into the superconducting phase and even the minimalistically designed doped RVB state considered in this work supports this feature, at least at the level of multiparty entanglement. Based on the behavior of both doped RVB states and exact GS of the $t$-$J$ Hamiltonian, one can hypothesize that the trend of $\cal{G}$ can detect the superconducting phase boundary, irrespective of the size of the ladder.
Note, however that we do not claim to detect a high $T_c$ superconducting phase by using the genuine multipartite entanglement  as an order parameter. This is also not the primary motivation of our work, which is to construct an efficient recursive method for evaluating bipartite as well as multiparty observables in large doped RVB states. As an useful spin-off to our main results, we are able to show that
for some parameter ranges, $\mathcal{G}$ may serve as an indicator to whether the system has entered into the SC phase or not. It is plausible that one would require further physical properties along with the GGM to identify all the phases. Since there exists, as yet,  no order parameter that can uniquely identify all the relevant phases of the ground states of doped Hubbard or $t$-$J$ model~\cite{sachdev}, the applicability of
$\mathcal{G}$ as a suitable order parameter requires further investigation. 

An important point in our work is the use of a non-variational RVB state as the GS of the doped quantum spin ladder.  It is clear that a variational RVB (vRVB) state, which lends possible support to $d$-wave pairing, is a more suitable state to study the doped ladder. However, vRVB states, in general, do not possess a recursive form that allows computation of reduced states with high efficiency in large systems, as the number of parameters to optimize increases exponentially. However, our results show that by omitting the variation in the coefficients of the covering required to build the  RVB state, we obtain a significant advantage in computation power, which allows us to compute $\cal{G}$ in large doped ladders. Comparison with exact GS of the $t-J$ ladder shows that the non-variational RVB state, quite accurately simulates the behavior of $\cal{G}$.
But we do think that efficient recursions for the certain variational RVB states  is an important problem for future tasks.

\begin{figure}[h]
\includegraphics[angle=0,width=6.0cm]{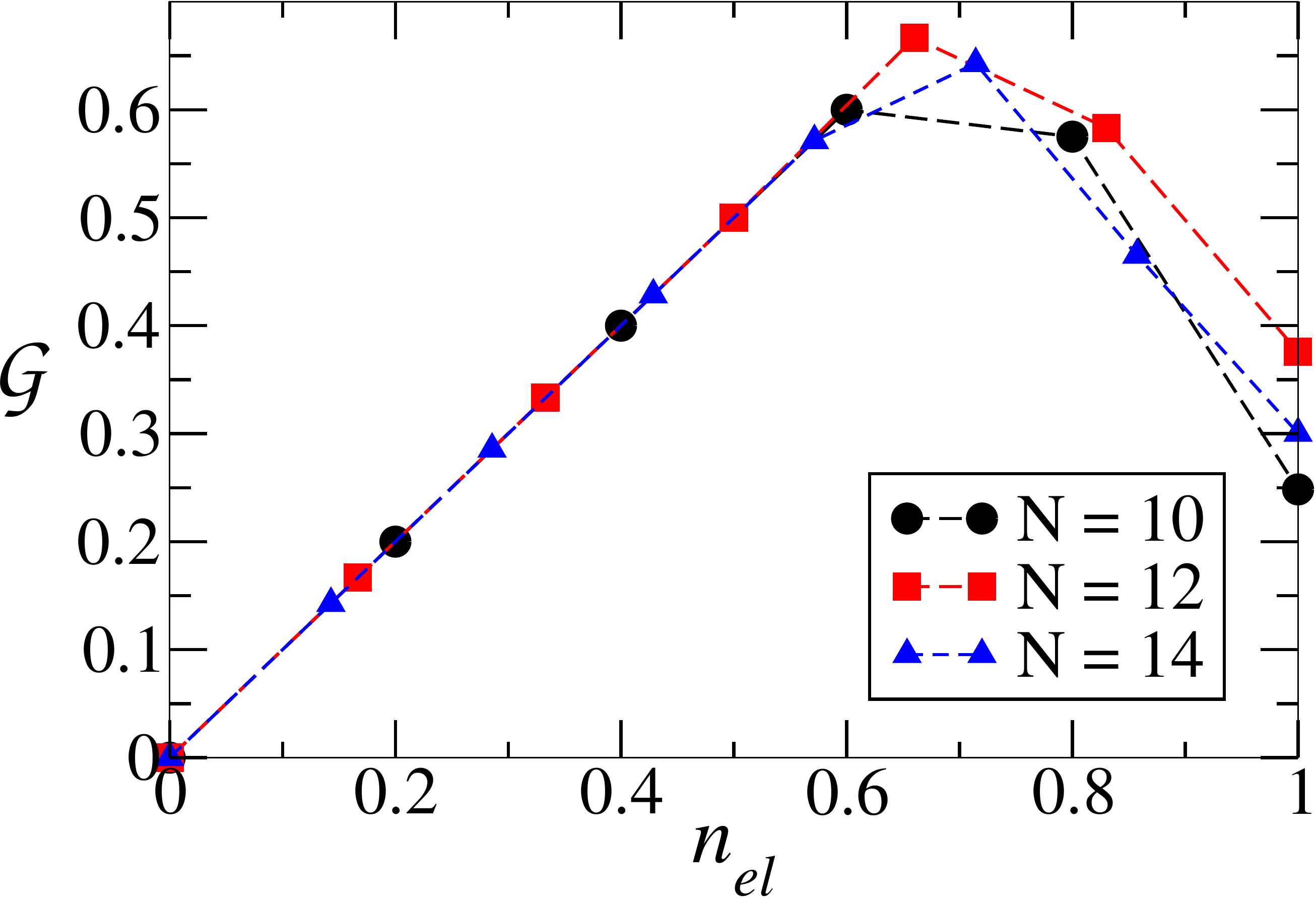}
\caption{(Color online). Genuine multisite entanglement in the $t$-$J$ ladder. Variation of $\mathcal{G}$ with $n_{el}$ for the exact GS of the $t$-$J$ ladder Hamiltonian, given in Eq.~(\ref{Ham_t_J}), for $N$ = 10, 12, and 14. $\mathcal{G}$ reaches its maximum value at $n_c \approx$ 0.65. Here $J/t=0.66$. All quantities plotted are dimensionless.}
\label{fig1}
\end{figure}
\vspace{-.5cm}
%%\begin{figure*}[t]
%%
%% \includegraphics[angle=0,width=6.2cm]{fig1.png}
%% \hspace{2mm}
%%\includegraphics[angle=0,width=9.2cm]{recursion.png}
%%
%% \caption{Schematic diagram of different parts of a 2-legged m-runged ladder (left). R.h.s. of the figure shows formation of $|\psi\rangle_{(3)',(1)}$
%% from $|\psi\rangle_{(4)',(0)}$ after introduction of a singlet.}
%%    \label{fig2}
%%  \end{figure*}
%{\it Recursion method for doped RVB states}:

\section{CONCLUSIONS}
\label{conclusion}
In this paper, we adopted two techniques for studying multisite entanglement   in doped quantum spin ladders. Firstly, we consider the doped RVB states as possible ground states of the $t$-$J$ Hamiltonian, which  we have shown to be always genuinely multiparty entangled. To overcome the limitations of exact diagonalization, we introduced a recursion method to generate the doped RVB state and to compute its reduced density matrices. By using the iterative method, we find that we can compute the genuine multiparty entanglement of doped RVB ladders, for large systems under finite doping of the ladder. We found that the maximum value occurs at doping concentration $n_{el}$ = 0.56. Secondly, we use an exact diagonalization method for the $t$-$J$ Hamiltonian, for upto 14 sites and observe that the GS of the Hamiltonian is also genuinely multipartite entangled, with maximum entanglement occurring at the superconducting phase boundary, where the electron density $n_{el} \approx$ 0.65.
 
We note that the primary outcome of our work is an analytical recursion method to evaluate the genuine multipartite entanglement in RVB ladders with finite doping. An immediate offshoot of our results is the connection between maximal entanglement in doped RVB states and the high-$T_c$ superconducting phases of the
$t$-$J$ Hamiltonian. In this regard, we would like to mention that even though GGM may apparently be useful in signalling  the onset of the high-$T_c$ superconducting phase, it is possible that the different phases of the Hubbard or the $t$-$J$ model can not be completely characterized   by just using entanglement. Recently, using the behavior of multiparty entanglement, attempts have been made to get more accurate insight about the phase boundaries which emerge in the ground state configuration of  $XXZ$ quantum spin ladders~\cite{xxz_zanardi,xxz_ours}. However, it has been shown that there are regions in the parameter space at which multiparty entanglement alone fails to a provide a conclusive phase diagram  and one needs to study the behavior of other ground state properties such as magnetization and
spin correlation functions in order to  obtain a complete picture of the different phase boundaries~\cite{xxz_ours}. In a similar vein, for the case of doped ladders, it is plausible that one would require further physical properties along with the GGM, to characterize different phases of the $t$-$J$ Hamiltonian. This requires further investigations on the model which are planned in forthcoming works. Apart from this, there has also been attempt to quantify the bipartite entanglement of certain multipartite pure states, like the Bardeen-Cooper-Schrieffer (BCS) state of superconducting compounds~\cite{delgado_BCS} which shed light on the relation of entanglement to that of the superconducting order parameter. We believe that the extension of such investigations to the case of high-$T_c$ cuprates may uncover interesting underlying microscopic properties.

%}
%
%
%our comments on superconductivity is a rather spin-off from the primary goal of this work, which aims to construct an efficient recursive method for evaluating multiparty observables in a large doped RVB state. However, we think that understanding the role of $\cal{G}$ in the variational RVB state, which also supports d-wave mechanism, is an important problem.
%We believe that our results are an addition to the broader understanding of the nature of quantum correlations in the states commonly associated with exotic superconductivity in strongly correlated system.

%Our results show that the superconducting states of the doped quantum spin ladders are maximally genuine multipartite entangled, at relatively high doping concentration, as described by the phase diagram of the $t$-$J$ Hamiltonian \cite{t-J_phase}. In this regime, one can expect that the strong correlation in the  superconducting currents should emanate from robust multiparty entanglement in the doped system although we note that there is no universal agreement on the phase separation or ground state correlations associated with high-$T_c$ superconductors. We conclude about such a high multiparty entanglement in a superconducting phase of a $t$-$J$ model by using exact diagonalization and by using doped RVB ansatz. For numerical exact diagonalization we choose Lanczos algorithm while to compute doped RVB states, we developed an analytical recursion method  capable of calculating any physical quantities of a state with a large number of sites.

\section{acknowledgement}
HSD acknowledges funding by the Austrian Science Fund (FWF), project no. M 2022-N27, under the Lise Meitner programme of the FWF.


\begin{thebibliography}{10}
\bibitem{comm2} We develop a numerical algorithm, based on the Lanczos method \cite{lanc}, to exactly solve the composite hole-dimer qutrit system. By dividing the Hilbert space in different subspaces, according to the hole concentration and total $S_z$, exact ground state of the $t$-$J$ Hamiltonian can be obtained for moderate system size. The algorithm is implemented using codes written in MatLab and Fortran90. Also, see the Appendix for more details.
\end{thebibliography}
\end{document}